# Seismic Facies Analysis: A Deep Domain Adaptation Approach

M Quamer Nasim, Tannistha Maiti, Ayush Srivastava, Tarry Singh, *Member, IEEE*, and Jie Mei

***Abstract*—Deep neural networks (DNNs) can learn accurately from large quantities of labeled input data, but often fail to do so when labelled data are scarce. DNNs sometimes fail to generalize ontest data sampled from different input distributions. Unsupervised Deep Domain Adaptation (DDA) techniques have been proven useful when no labels are available, and when distribution shifts are observed in the target domain (TD). In the present study, experiments are performed on seismic images of the F3 block 3D dataset from offshore Netherlands (source domain; SD) and Penobscot 3D survey data from Canada (target domain; TD). Three geological classes from SD and TD that have similar reflection patterns are considered. A deep neural network architecture named EarthAdaptNet (EAN) is proposed to semantically segment the seismic images when few classes have data scarcity, and we use a transposed residual unit to replace the traditional dilated convolution in the decoder block. The EAN achieved a pixel-level accuracy >84% and an accuracy of ~70% for the minority classes, showing improved performance compared to existing architectures. In addition, we introduce the CORAL (Correlation Alignment) method to the EAN to create an unsupervised deep domain adaptation network (EAN-DDA) for the classification of seismic reflections from F3 and Penobscot, to demonstrate possible approaches when labelled data are unavailable. Maximum class accuracy achieved was ~99% for class 2 of Penobscot, with an overall accuracy>50%. Taken together, the EAN-DDA has the potential to classify target domain seismic facies classes with high accuracy.***

## I. Introduction

INTERPRETATION of geologic features and inference of reservoir properties are key to the success of hydrocarbon exploration and production. Accurate delineation of subsurface structures is a necessary and routine process in seismic interpretation. Automation of this task will allow for timely delivery of interpreted seismic sections to support prospective zone identification, well planning, reservoir modeling, and geohazard analysis. In recent years, there is a massive interest in the application of DNNs for automating seismic interpretation [2-9].

Unfortunately, large publicly available annotated datasets for seismic interpretation are sparse, making the application of traditional deep learning methods challenging. To overcome this challenge, researchers often annotate their own training and testing datasets which is a time consuming process [2]. Few options to overcome scarcity in annotated data include a) weakly-supervised learning approaches [10], b) similarity-based data retrieval [1], and, c) weakly-supervised label mapping algorithm. Studies have also used unsupervised machine learning techniques, such as principal component analysis or self-organizing maps [11-13]. Alternatively, researchers have proposed new architectures like Danet-FCN2, and Danet-FCN3 [14] that replaced the traditional dilated convolutions in the decoder block with a transposed residual unit thus reduced the amount of training data required.

The use of transfer learning with an already trained DNN can significantly reduce the costs associated with model training from scratch and leads to a high classification accuracy even with a smaller amount of training data [15]. To effectively apply knowledge acquired from one task to a different task in semantic classification, [16]used transfer learning and showed that a DNN trained with one seismic dataset could be reused in another similar task i.e., seismic facies semantic classification. [17] showed that trained DNN models would under-perform when tested on samples from a related, but non-identical domain by using transfer learning. However, transfer learning is still challenging in areas such as medical imaging and earth science because large annotated datasets are required for the models to benefit from the inductive transfer processes [18, 19].

Compared with natural image datasets, DDA for cross-modality images in earth science is more challenging. The existence of the domain shift is common in real-world applications [20, 21], where the semantic class labels are usually shared between domains while the distributions of data are different. For example, seismic images are acquired in different stratigraphic settings and are related by reflection patterns, stratigraphic settings, and depositional environments. These images differed due to different stratigraphic settings and depositional environments. Distributions of these data mismatch significantly given their different density, porosity, rock types, and permeability in the Earth (Fig. 1).

To address this issue, unsupervised DDA methods have been proposed and evaluated to allow generalization of the trained models to new datasets [22]. The domain of labeled training data is termed as the SD, and the test dataset is called the TD.

M Quamer Nasim and Ayush Srivastava is with the Indian Institute of Technology, Kharagpur, West Bengal, 721302, India and also with deepkapha.ai, Amsterdam, Netherlands (e-mail: quamer23nasim@yahoo.com; quamer.nasim@deepkapha.com; ayush.srivastava@deepkapha.ai).

Tannistha Maiti was with University of Calgary. She is now with deepkapha.ai, Amsterdam, Netherlands (e-mail: tannistha.maiti@deepkapha.ai).

Tarry Singh is with deepkapha.ai, Amsterdam, Netherlands (e-mail: tarry.singh@deepkapha.ai).

Jie Mei is with University of Quebec at Trois-Rivieresand also with deepkapha.ai, Amsterdam, Netherlands (e-mail: jie.mei@deepkapha.ai; jie.mei@uqtr.ca).

GitHub: https://github.com/deepkapha/EarthAdaptNet

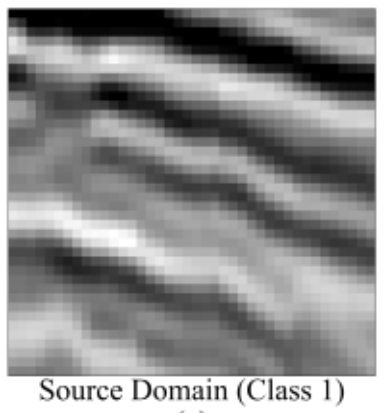

Source Domain (Class 1)
(a)

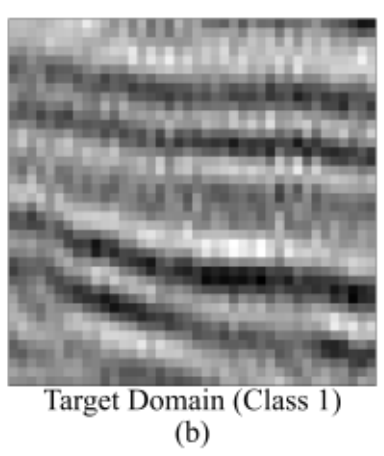

Target Domain (Class 1)
(b)

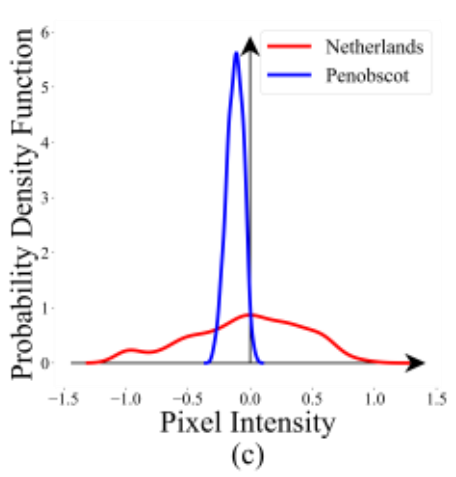

Fig. 1. Domain Shift exists in seismic reflection pattern in terms of intensity and resolution (a) Reflection pattern of Class 1 F3 Block, Netherlands i.e., Source Domain (b) Reflection pattern class 1 Penobscot, Canada i.e., Target Domain (c) Histogram of Pixel intensity

The Unsupervised DDA methods are more feasible over transfer learning, given that this methodology transfers knowledge across domains without the need for TD labels. To the best of our knowledge, study of effective generalization of trained DNNs across domains for seismic images has not been investigated yet. In this study, we focus on domain knowledge transfer between two different stratigraphic locations using reflection pattern similarities. A reflection pattern is a property of seismic reflectors, and in sequence stratigraphy, it refers to patterns observed in reflectors present in seismic sections, e.g., high amplitude reflectors, low amplitude reflectors, as well as parallel, subparallel, chaotic reflectors. Data distribution, on the other hand, refers to the property of data and defines the statistics of data irrespective of domain. Data distribution can be further categorized into gaussian (normal), uniform, and beta distributions.

In this article, we present an approach that exploits accurate and robust semantic segmentation (classification) of seismic images with cropped local image patches on the F3 block of the Netherlands. Our architecture especially focuses on classes that have scarce labeled data and leads to higher accuracy. We propose a network architecture with Residual Blocks (RBs) and Transposed Residual Blocks (TRBs) with skip connections between the two to address the issue of vanishing gradients. We also introduce the concept of DDA to bridge the gap between SD and TD in a joint space.

The main contributions of this article are as follows.

1) We propose a network architecture named EarthAdaptNet for accurate delineation of seismic facies which can achieve higher performance in comparison to the baseline architectures, especially for minority classes. Here, we apply EAN architecture to classify seismic facies of the F3 block.
2) We redesign EarthAdaptNet to incorporate CORAL (Correlation Alignment for Domain Adaptation) method by constructing a differentiable loss function that minimizes the difference between source and target correlations, i.e., the CORAL loss, which learns the non-linear transformation between source and target correlations. DDA method proves useful when labelled data are not available, which is quite often the case in seismic studies.
3) We assess performance of the proposed EAN-DDA in a multi-class classification problem to analyze seismic facies. The seismic facies dataset contains 3 representative facies classes that have similar depositional and compositional environments. We follow the standard protocol of domain adaptation [23] and use all labeled SD data and unlabeled TD data. We also generate patch images for the domain adaptation facies classification problem and publicly available.

The remainder of this article is organized as follows. A summary of network architectures used and the proposed approach can be found in Section 2. A detailed description of the background of DDA and applications of the CORAL method to EarthAdaptNet is provided in Section 3. A description of the datasets used in the semantic segmentation and steps to generate cropped patch images for DDA analysis is given in Section 4. Performance metrics used in this study are also defined in this section. Experimental results are presented and assessed in Section 5. Discussion on results of this study is given in Section 6. In the end, we conclude this study and propose future research directions in Section 7.

## II. Proposed Network Architectures

In the present study, we propose state-of-the-art architectures, i.e., EarthAdaptNet and its variants, for semantic segmentation of seismic facies. We also present the EAN-DDA architecture for DDA of seismic facies. The two approaches for training a neural network on seismic data of relatively large section sizes are:1) Direct model training on large seismic sections using section-based models, which often requires high computational power given the section size is $701X255$ $(Crossline\ X\ Depth)$ and $401X255$ $(Inline\ X\ Depth)$. 2) Model training using small patches created out of large seismic sections, which is relatively computationally inexpensive. Once a model is trained, while predicting on the test set, one can quickly regenerate the whole section from patches, thereby having seismic sections as the final output. A model trained by using this type of approach is called a patch-based model. While section-based models may yield higher accuracy since they can process the whole image in one go as opposed to patch-based models, which tend to utilize more computational resources that are not always available. There is no overlap of data in section-based models, while in patch-based models, to increase the number of training examples, patches tend to overlap. In this paper, we will be using the patch-based model for both studies.

### A. *EarthAdaptNet*

The architecture EarthAdaptNet (Fig. 2) proposed for semantic segmentation is inspired by U-Net [24] and Danet-FCN3[14, 25]. Originally proposed for biomedical image segmentation, U-Net uses a contracting path to capture context and symmetrical expanding paths for attaining the original size of the input. Contracting and expanding paths are accompanied by shortcut connections at each level. Danet-FCN3 uses RBs and TRBs for semantic segmentation of seismic images. We therefore use U-Net to combine low level and high-level features [14] and Danet-FCN3 to overcome the issue of vanishing gradients [25, 26]. Building blocks of EarthAdaptNet can be broadly divided into RBs [26] and TRBs [25] similar to those of Danet-FCN3 but with some modifications. In the proposed architecture, RB comprises two convolutional layers, each followed by batch normalization and a downsampling residual connection of a $1X1$ convolutional layer. In view of U-Net, this is what is referred to as the building block of the contracting path. TRB is similar in architecture to RB except with the use of a transposed convolutional layer instead of a convolutional layer. Upsampling transposed residual connection with a $1X1$ convolutional layer is used instead of downsampling residual connection. EarthAdaptNet uses Encoder-Decoder architecture with RBs and TRBs. The encoder starts with a convolutional layer and is followed by the RB, and the number of RBs used depends on the input size. In this study, we experimented with 3-5 RBs. The decoder starts with a TRB and the number of TRBs used is kept the same as the RB used in the Encoder. The transposed residual layer is followed by a transposed convolutional layer which outputs the segmented seismic image. A $1X1$ convolutional layer also exists in the middle which acts as a bridge (bottleneck) between the Encoder and the Decoder. Skip connection is present between each RB and TRB.

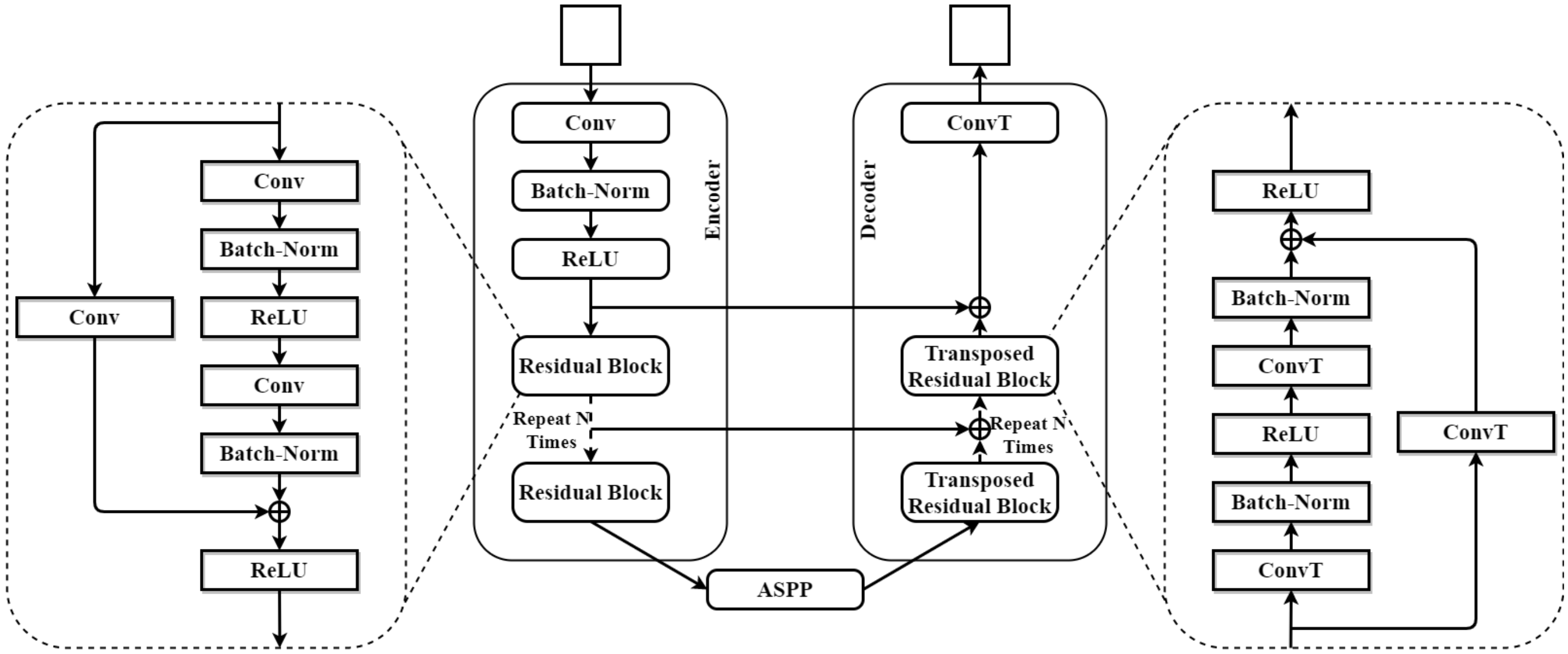

Fig. 2. Building blocks of the proposed EarthAdaptNet, which consists of RBs and TRBs and ASPP. RB comprises two convolutional (Conv) layers, each followed by batch normalization and a downsampling residual connection of the 1x1 Conv layer. TRB is similar in architecture as RB except with the use of a transposed convolutional (ConvT) layer instead of a convolutional layer. The encoder starts with a Conv Layer and is followed by the RB. Decoder starts with a TRB and the number of TRBs used is kept the same as the RB used in the Encoder and is followed by a Transposed Convolutional Layer which outputs the segmented seismic image. A ASPP module also exists in the middle which acts as a bridge (Bottleneck) between the Encoder and the Decoder (Fig. 3). Skip connection is present between each RBs and TRBs.

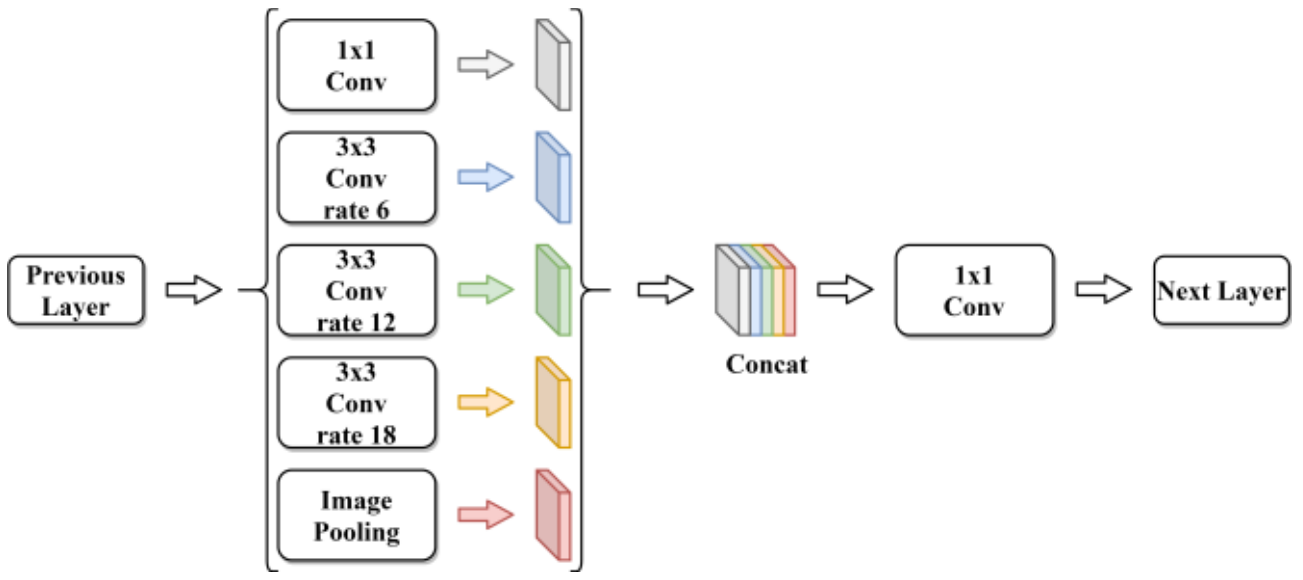

Fig. 3. Atrous spatial pyramid pooling (ASPP) module having 5 parallel layers with three layers being 3x3 convolutional layers with dilatation rate of 6, 12, and 18. One layer being 1x1 convolutional layer and last layer being image pooling layer. Output of all of these 5 layers is concatenated followed by final 1x1 convolutional layer.

We also introduced the atrous spatial pyramid pooling (ASPP) module from DeepLab V3 [27] in our EarthAdaptNet architecture in order to capture multi-scale features. ASPP module has 5 parallel layers, three of which are atrous convolutions of a $3X3$ filter size with different dilatation rates (i.e., 6, 12, and 18) and one $1X1$ convolution layer and lastly, an image pooling layer. Each parallel layer in the ASPP module has 256 filters followed by a batch normalization layer. Finally, output of all 5 parallel layers is concatenated followed by another $1X1$ convolution with 256 filters. With the help of 5 parallel layers and different atrous convolution rate, the ASPP module is designed to capture multi-scale information (Fig. 3).

An important point to note in the EAN architecture is that there is no batch normalization layer in the shortcut connection. Raw output of the convolutional layer from shortcut connection is added to batch normalization layer from the main path. We first trained our model with the batch normalization layer in the shortcut connection. However, performance of this model was very poor (Supplementary Table V) and didn't improve after hyperparameter tuning. After removing the batch normalization from the short connection, good performance with improved accuracy was obtained. The results presented in the subsequent chapters are based on the model which does not have batch normalization layer in shortcut connection.

Experiments were performed to examine the following architectures:

1) EarthAdaptNet model with a middle convolutional layer in 4 RB-TRB pairs;
2) EarthAdaptNet model with a middle convolutional layer in 5 RB-TRB pairs;
3) EarthAdaptNet model with a ASPP module in place of the middle convolutional layer in 4 RB-TRB pairs.

### *B. EarthAdaptNet Deep Domain Adaptation network (EAN-DDA)*

We revisited the EarthAdaptNet (EAN; Fig. 4) and created three variations of this architecture for DDA study, including:

1) 4 RBs followed by 4 fully connected layers (*4RB + 4 FC*; Table I);

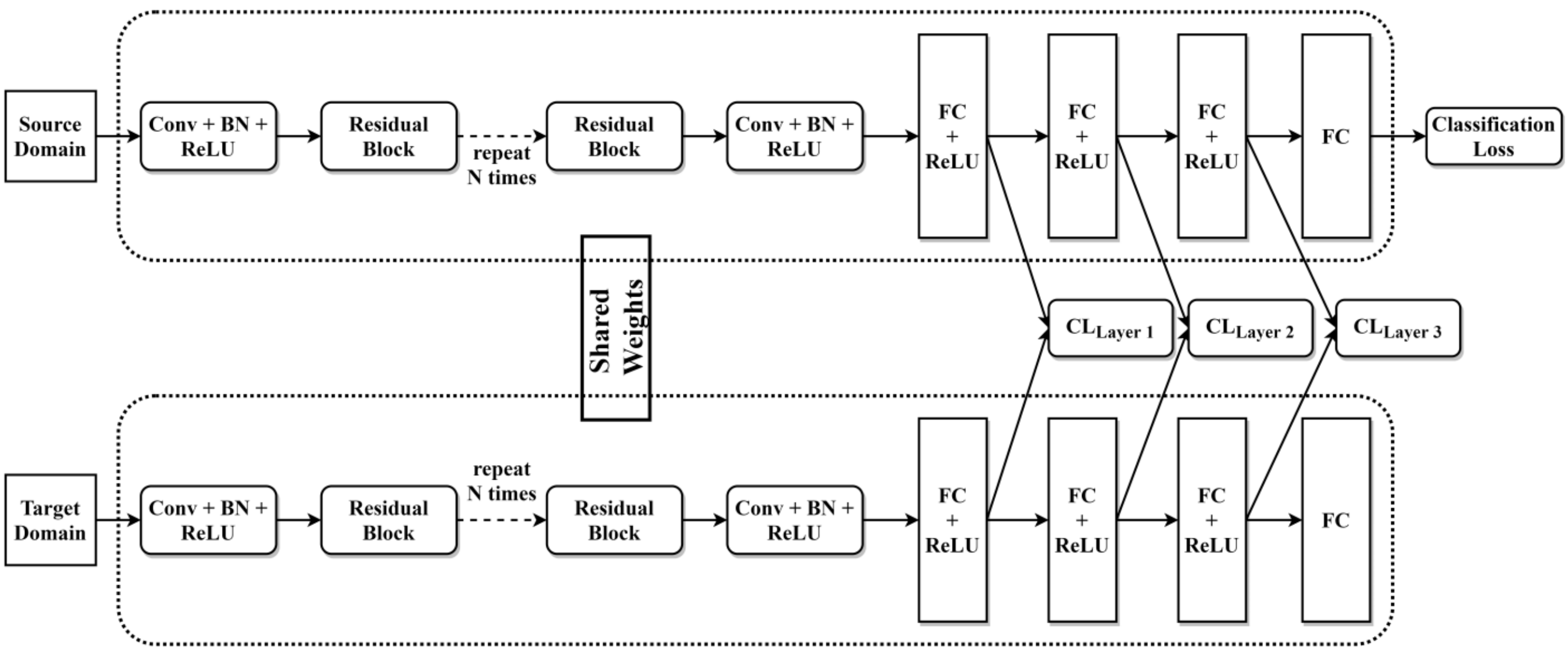

Fig. 4. DeepCORAL architecture with the backbone of EarthAdaptNet's encoder which consists of RBs only. RBs are shown in Fig 2. We apply the CORAL loss to the three FC + ReLU layers and apply Classification loss to the last layer of SD network. Source: F3 cropped patch images, Target: Penobscot cropped patch images.SD and TD network shares common weights across the architecture

TABLE I
UNITS LAYERWISE ARCHITECTURAL DESCRIPTION OF THE FEATURE MAPS FOR THE SOURCE AND TARGET DOMAINS FOR 4 ENCODERS FOLLOWED BY 4 FC LAYERS (*4RB+4FC*)

| Entity | Entity Size | Entity Description |
|---|---|---|
| *input_source* | (128, 1, 40, 40) | SD image batch |
| *input_target* | (128, 1, 40, 40) | TD image batch |
| *first_conv_source* | (128, 64, 40, 40) | SD image batch feature map after first conv layer |
| *first_conv _target* | (128, 64, 40, 40) | TD image batch feature map after first conv layer |
| *RB_1_source* | (128, 64, 20, 20) | SD image batch feature map after first RB |
| *RB_1_target* | (128, 64, 20, 20) | TD image batch feature map after first RB |
| *RB_2_source* | (128, 128, 10, 10) | SD image batch feature map after second RB |
| *RB_2_target* | (128, 128, 10, 10) | TD image batch feature map after second RB |
| *RB_3_source* | (128, 256, 5, 5) | SD image batch feature map after third RB |
| *RB_3_target* | (128, 256, 5, 5) | TD image batch feature map after third RB |
| *RB_4_source* | (128, 512, 3, 3) | SD image batch feature map after fourth RB |
| *RB_4_target* | (128, 512, 3, 3) | TD image batch feature map after fourth RB |
| *middle_conv_source* | (128, 512, 3, 3) | SD image batch feature map after last conv layer |
| *middle_conv_target* | (128, 512, 3, 3) | TD image batch feature map after last conv layer |
| *FC1_source* | (128, 2048) | SD image batch feature map after first FC layer |
| *FC1_target* | (128, 2048) | TD image batch feature map after first FC layer |
| *FC2_source* | (128, 1024) | SD image batch feature map after second FC layer |
| *FC2_target* | (128, 1024) | TD image batch feature map after second FC layer |
| *FC3_source* | (128, 512) | SD image batch feature map after third FC layer |
| *FC3_target* | (128, 512) | TD image batch feature map after third FC layer |
| *FC4_source* | (128, 3) | SD image batch feature map after last FC layer |
| *FC4_target* | (128, 3) | TD image batch feature map after last FC layer |

2) 4 RBs followed by Global Average Pooling (GAP) layer followed by 2 fully connected layers (*4RB + GAP + 2FC*; Supplementary Table I);
3) 3 RBs followed by Global Average Pooling layer followed by 2 fully connected layers (*3RB + GAP + 2FC*; Supplementary Table II).

Table I, Supplementary Table I and Supplementary Table II show all the components for the respective architecture along with the size of the components.

The initial distribution of a random sample (batch size 128) is shown in Fig. 9. As evident from the distribution plots based on [28], the source and target distributions differ greatly. Due to difference in data distribution, the performance of traditional deep learning approaches is compromised. Therefore, we introduced the EAN-DDA architecture based on DDA methodology which essentially bridges the gap between SD and TD. We discuss three variations of the EAN-DDA network with individually fine-tuned hyperparameters.

For the first variant of EAN-DDA model (*4RB + 4FC*), we applied CORAL loss to all fully connected (FC) layers except the final output layer while for second (*4RB + GAP + 2FC*) and third (*3RB + GAP + 2FC*) variant, we applied GAP to flatten the results from the last encoder layer and then use CORAL loss to GAP layer and the 1st FC layer. It's a common practice to use GAP layer since it's a fairly simple operation that reduces the data significantly and prepares the model for the final classification module. We used CORAL loss with a weighting factor. Weighting factors in first variant (*4RB + 4FC*)are 0.2, 0.6, and 0.2 for 1st, 2nd, and 3rdFC layers, respectively while for second (*4RB + GAP + 2FC*) and third (*3RB + GAP + 2FC*)variant weighting factors are 0.5 and 0.5 for the GAP and the 1st FC layers, respectively. In the third variant of EAN-DDA model (3RB + GAP + 2FC) we decreased the number of RBs from 4 to 3 to see the model's performance, given that we're dealing with a patch size of $40X40$ only.

We initialized the network parameters from a pre-trained network (Non-DDA Model) and fine-tuned it using the labeled SD data. The dimension of the last fully connected layer was set to the number of classes (i.e., 3), with weights initialized with $N(0, 0.005)$ [23]. The learning rate of the last fully connected layer is set to 10 times the other layers as it was trained from scratch. The weight of the CORAL loss (λ) is set in a way that at the end of the training the classification loss and CORAL loss are roughly the same [23].

## III. Background on Domain Adaptation - EarthAdaptNet Unsupervised Domain Adaptation

In this study, we propose DDA method using the DeepCORAL (Correlation Alignment) [23] methodology for seismic facies classification, which uses CORAL Loss [23] to match the data distribution of the SD and TD at various feature layers. For this purpose, we first introduce CORAL loss for a single feature layer. Let us assume that we have SD and TD $d$-dimensional encoded features from a particular feature layer as $D_S$and $D_T$, and the total amount of SD and TD data samples are $n_S$ and $n_T$, respectively. $D_S^{ij}$ $(D_T^{ij})$represents the $j^{th}$ dimension of the $i^{th}$ SD (TD) encoded feature for a particular feature layer and $C_S(C_T)$ denotes the feature covariance matrices. CORAL loss is defined as the distance between the second order statistics (covariances) of the SD and TD encoded features [23]:

$$l_{CORAL} = \frac{1}{4d^2} \|C_S - C_T\|_F^2 \tag{1}$$

Where $\|.\|_F^2$ represents the squared matrix Frobenius norm and can be calculated as follows. The covariance matrices of the SD and TD data are given by:

$$C_S = \frac{1}{n_S - 1}\left(D_S^T D_S - \frac{1}{n_S}(\mathbf{1}^T D_S)^T(\mathbf{1}^T D_S)\right) \tag{2}$$

$$C_T = \frac{1}{n_T - 1}\left(D_T^T D_T - \frac{1}{n_T}(\mathbf{1}^T D_T)^T(\mathbf{1}^T D_T)\right) \tag{3}$$

$$\|C_S - C_T\|_F^2 = trace\big((C_S - C_T)^*(C_S - C_T)\big) \tag{4}$$

Where$\mathbf{1}$in $\mathbf{1}^T$is a column vector with all elements equal to 1, which should not be confused with an identity matrix, and $(C_S - C_T)^*$is a conjugate transpose which can be computed as:

$$(C_S - C_T)^* = (\overline{C_S - C_T})^T \tag{5}$$

The gradient with respect to the input features can be calculated as follows:

$$\frac{\partial l_{CORAL}}{\partial D_S^{ij}} = \frac{1}{d^2(n_s - 1)}\left(\left(D_S^T - \frac{1}{n_S}(\mathbf{1}^T D_S)^T \mathbf{1}^T\right)^T (C_S - C_T)\right)^{ij}$$

$$\frac{\partial l_{CORAL}}{\partial D_T^{ij}} = \frac{1}{d^2(n_T - 1)}\left(\left(D_T^T - \frac{1}{n_T}(\mathbf{1}^T D_T)^T \mathbf{1}^T\right)^T (C_S - C_T)\right)^{ij}$$

For the classification of seismic classes, we used a cross-entropy loss function. The CORAL loss is extended to total$t$ feature layers. By training the data on both classification and CORAL loss features are learned that work well on target domain.

$$l = l_{CLASSIFICATION} + \sum_{i=1}^{t} \lambda_i l_{CORAL} \tag{6}$$

Where$t$ denotes the number of CORAL loss layers in a deep neural network and λ represents the weight on each CORAL loss applied to $t$ encoded feature layers. A difference between the ranges of classification loss and of CORAL loss was observed and a normalization factor is used to bring the two losses to comparable ranges. An important point regarding the EAN-DDA study is that the classification error is calculated for SD, for which ground truth of SD is required. In contrast, the CORAL loss is calculated between SD and TD, which does not require ground truth. Hence, in the EAN-DDA study, we only need ground truth for SD.

For this study, we took two datasets with different distributions, i.e., SD and TD. SD had labels while TD didn't. We then defined the CORAL loss as per equation 1 by first obtaining the covariance matrix of SD and TD, then calculated the Frobenius norm between the covariance matrices of SD and TD. The EAN-DDA network is composed of two parts, first being a feature extractor of Convolutional Layers, and a Classifier of FC Layers. We initiated the two parallel networks using shared weights (Figure 4) for SD and TD, respectively. In the SD network, we applied cross entropy loss in the output layer, and we applied CORAL loss to encoded features from all FC Layers except the output layer. We sum up the CORAL Loss with cross entropy loss, and use the final resulting loss in backpropagation and optimize the model parameters via the Adam Optimizer. Once the model is trained, we have a model that can map SD and TD to a distribution-invariant feature map which then is utilized to predict labels on TD using only the SD network.

## IV. Seismic Facies Dataset

### A. Dataset

This study uses processed seismic data collected from the F3 block in the Netherlands and Penobscot in Canada. Generating seismic images is a sophisticated process that involves data acquisition, where intense sound sources are placed between 6 and 76 m below the ground to generate sound waves. These waves pass through different layers of rock (strata) and are

TABLE II

INTERPRETED HORIZONS OF NETHERLANDS F3 BLOCK AND CANADA PENOBSCOT DATA. COMPARISONS ARE MADE BASED ON COMPOSITION AND DEPOSITIONAL ENVIRONMENT AND REFLECTION PATTERNS. REPRESENTATIVE CLASSES ARE THE CLASS NAMES USED IN DA STUDIES

| Formation | Compositional and Depositional Environment | Reflection Pattern | Representative Class |
|---|---|---|---|
| Chalk and Rijnland | Clay Formations with Sandstones; Coastal shallow to fairly deep open Marine environment | Parallel and High-Amplitude Reflectors | Class 1 |
| Scuff | Claystones; Shallow Marine to continental Marine environment from restricted(lagoonal) to open Marine (outer shelf) condition | Subparallel and Varying-Amplitude Reflectors | Class 2 |
| Zechstein | Evaporites and Carbonates; Peri Marine to Marine settings | Continuous and Low-Amplitude Reflectors | Class 3 |
| H6-H5 | Carbonates and Clastics of Iroquois Formation and Coarse Clastic Fluvial sediments of Mohican Formation; Shallow Marine setting | Parallel, High-Amplitude, and Chaotic Reflectors | Class 1 |
| H5-H4 | Fine-grained Glaciomarine, Gravel, 3-41% Sand, 30-56% Silt, and 29-45% Clay; Marine transgression, Reflectors having Prograding Sigmoidal Configuration of Low Energy and Medium to Low Amplitude due To Complex Delta System Deposition | Subparallel and Varying-Amplitude Reflectors | Class 2 |
| H4-H3 | Coastal area having a complex history of glaciation and sea-level rise | Continuous and Low-Amplitude Reflectors | Class 3 |

reflected, returning to the surface, where geophones or hydrophones can record them. This signal is then processed using an iterative procedure to generate seismic images. Finally, interpreters analyze the generated images and divide them into the different categories, or facies [29]. These categories represent the overall seismic characteristics rock unit that reflects its origin, differentiating this unit from the other ones around it [29]. It consists of a horizontal stack of 2D seismic images (slices), leading to a 3D volume, with the vertical axis of this volume representing its depth. The remaining axes define the inline and crossline directions. Geoscientists based their interpretations of facies based on configuration patterns that indicate geological factors like lithology, stratification, depositional systems, etc. [30].

In this work, we use a publicly available fully annotated dataset from the Netherlands F3 Block(https://github.com/olivesgatech/facies_classification_benchmark). The inline slices are the images in the cube perpendicular to the inline direction. The same idea applies to the crossline slices, which are images along the depth axis and perpendicular to the crossline axis. The F3 dataset included 401 crossline and 701 inline slices, with a dimension of$401X701$. In a previous study [2], the slices were interpreted and annotated, and a label mask was generated for each slice. F3 block seismic data consist of sections from inline 100 to 701 and crossline 300 to 1201. The whole dataset was divided into 3 smaller subsets, namely Train, Test #1 and Test #2. To avoid data leakage and overestimation of model performance, no overlap exists between the training and testing sets [42, 43]. The main dataset included all data in the ranges of inlines [100,700] and crosslines [300, 1200]. The training set contained the sections in the ranges of inline [300, 700] and crossline [300, 1000], Test set #1 contained sections in the ranges of inline [100, 299] and crossline [300, 1000], and Test set #2 contained sections in the ranges of inline [100, 700] and crossline [1001, 1200] [2]. The main lithostratigraphic unit of the F3 block are the Upper North Sea group, the Lower and Middle North Sea group, Chalk group, Rijnland group, Schieland, Scruff and Niedersachsen groups, Altena group, Lower and Upper Germanic Trias groups, and Zechstein group, arranged according to their depth, with the Upper North Sea group representing shallowest horizon and Zechstein group representing deepest horizon.

We used another publicly available fully annotated dataset of Penobscot, Canada (https://zenodo.org/record/1324463#.X5cwfFgzbIU).The dataset was used as target domain for training in DDA studies. However, during the training of EAN-DDA model the annotated version with target labels were not used, and only annotations for validation set was used. The Penobscot dataset included 481 crossline slices and 601 inline slices, with dimensions $601X1501$ and $481X1501$ pixels, respectively. Although the unsupervised DDA study doesn't require splitting the dataset, as we tried to perform hyperparameter tuning and to minimize the bias introduced, we split the dataset into a training set and a test set. The training set includes data from the ranges of inline [1000, 1500] and crosslines [1000, 1480], while the test set included data from the ranges of inlines [1500, 1600] and crosslines [1000, 1480] [31]. All slices have been interpreted and annotated, and a label mask was generated per slice. The seven interpreted horizons: H1, H2, H3, H4, H5, H6, and H7 [31] are numbered according to their depth, with H1 representing the deepest horizon and H7 representing the shallowest horizon.

### B. *Representative Facies classes*

Seismic stratigraphy [32], in conjunction with sequence stratigraphy [33], are two interpretation techniques developed to help predicting facies and reservoir distribution that add the time dimension to the depositional models [34]. Sequence stratigraphy is a complex model which is essentially based on sea-level changes and seismic stratigraphy is a technique that facilitates stratigraphic interpretation of seismic reflectors. Essentially, sequence stratigraphy applies the geological concepts of stratigraphy to the interpretation of seismic data. The basic assumption behind seismic stratigraphy is that individual reflectors can be considered as timelines, i.e., each represents a very short time interval of similar sedimentation conditions. This assumption signifies that a seismic reflector formed at different depositional environments and therefore it contains information of various lithofacies units. Seismic facies are classified based on reflection patterns including reflection configuration, reflection continuity, reflection amplitude and reflection terminations, based on which several geological interpretations could be made.

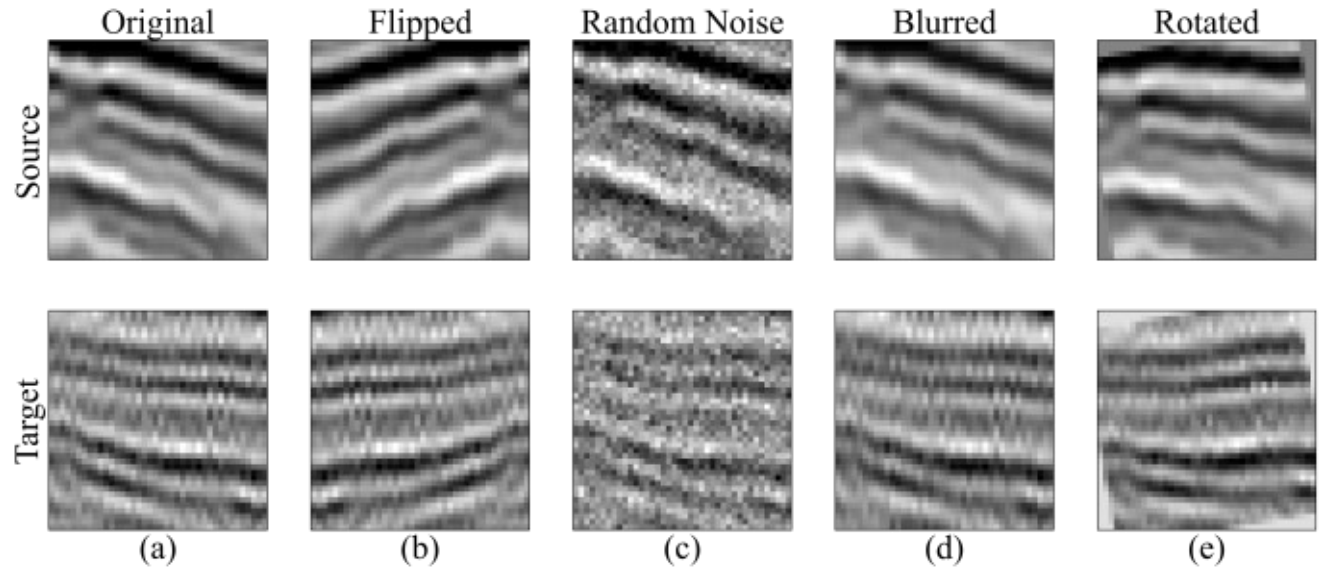

Fig. 5. Sample augmentation to generate cropped patch images to solve DA classification problem; Source: F3 Block Target: Penobscot block. Original samples in (a) are flipped (b) random noise added, (c) blurred, and (d) rotated.

In this study, we have selected three representative facies classes from the F3 Netherlands block [2] and Penobscot data[35]. The representative facies were selected based on their lithofacies composition, depositional environment and reflection patterns (Table II). The comparable facies classes from F3 and Penobscot are renamed as Class 1, Class 2 and Class 3 respectively based on similar reflection pattern and is used for comparison in DA studies. The depositional environment for the F3 block is predominantly shallow marine-to-marine depositions along with restricted marine and floodplain settings, while the Penobscot block has marine water filled basins. We have shown the domain shift between SD and TD in Fig. 6. Though reflection patterns seem similar, there is a wide variation in characteristics (Table II) which causes significant domain shift. An important point to note in the DDA dataset is that the reflection patterns of Chalk and Rijnland of the Netherlands and H6-H5 of Canada are slightly different in terms of amplitude and continuity. This difference might lead to lower accuracy as compared to the other two classes.

### C. *Generate patches from the dataset*

The patch-based model for segmentation problem extracts 2-D patches of a given size from the seismic sections, i.e., inline and crossline sections which itself is extracted from seismic volume along with their masks [2]. We have used patches of a dimension of $99X99$. The stride is set to half the size of the patch. A window of a given size (patch size) moves all over a section, whose$i^{th}$ pixel is apart from the $i^{th}$ pixel of adjacent window by half the value of given patch size. Once all the patches were extracted using the above-mentioned method, 20% of them were kept aside to use for validation set. We also reconstructed seismic sections from seismic patches for the evaluation of model performance in the testing phase. It should be noted that during training, the input consists of overlapped patches from seismic sections, while in the training process; we didn’t regenerate the sections from patches. We applied loss function to the output of patches itself, and when testing the model, we only used non-overlapping patches and then regenerated seismic sections out of non-overlapping patches hence there was no need to aggregate every patch output and average them.

In the DDA study, one requires a slightly different kind of dataset as compared to classical machine learning approach. Instead of having just one dataset, in DDA study we need two different datasets, i.e., datasets from both Source Domain (SD)and Target Domain (SD) and the two datasets should have different data distributions while the task of the model remains the same. As publicly available geophysics datasets do not fulfill this requirement, instead of predicting class with similar facies, we would like to predict similar reflection patterns (Table II).

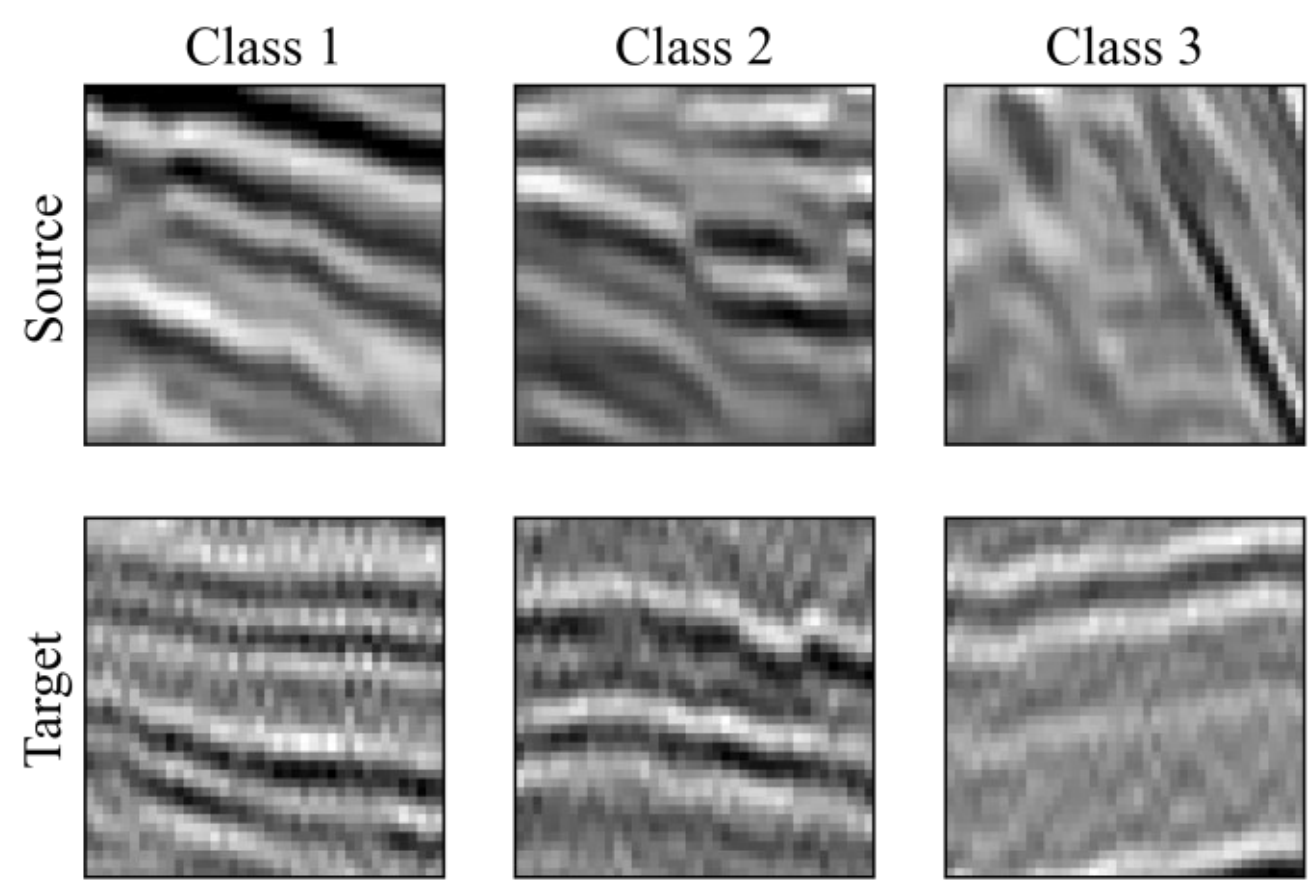

Fig. 6. All representative classes i.e., SD-TD pairs used in EAN-DDA study. Top row represents SD and bottom row represents TD. While each column represents a representative class of SD and TD. It can be clearly seen that though the reflection pattern of representative class of SD and TD are same, yet their appearances are different hence proving that there is a domain shift between datasets. Reflection pattern is a property of seismic facies while data distribution is a property of any kind of data, which defines the statistics of the data.

To approach the classification problem in DDA study, we generated a patch size of $40X40$ from both SD (the Netherlands) and TD (Canada). Since the study of Domain Adaptation in seismic reflection patterns has not yet been performed by other researchers, we created our own dataset for the study. We defined a representative class as a class that has similar reflection patterns in SD with TD (Table II). Given the differences in geological formations of seismic facies of F3 Block and Penobscot, we performed the DDA study based on their reflection patterns. Out of 6 classes in F3 Block [3] and 7 classes in Penobscot [31], only 3 classes had similar reflection patterns (Table II). To generate the dataset as per the DDA requirements discussed above, we removed patches of different reflection patterns within SD and TD (3 classes for F3 Block and 4 classes for Penobscot). After that, a single-valued label indicating the representative class is then assigned to each patch, and if 70% or more pixels of a patch belong to a particular class, we assign the corresponding single-valued class label to that patch. If less than 70% of all pixels of a patch belong to a particular class, the patch was excluded. Taken together, we formulated the EAN-DDA study as a classification problem instead of a segmentation problem.

### D. *Data Augmentation*

Data augmentation enables practitioners to significantly increase the diversity of data available for training deep learning models without the need to collect new data. Fig. 5 illustrates the augmentations applied to the cropped patches, including random rotation (≤10 degrees), blurring, flipping, shifting, and adding random noises [2]. Previous studies [2]showed that data augmentation significantly improved the performance of both baseline models, i.e., patch-based model and section based model , but the effect was more pronounced for the patch-based model. [2] found that the frequency weighted intersection over union (FWIoU) and mean class accuracy (MCA) scores increased by more than 10% in the patch-based model, while model performance was improved more significantly on smaller classes such as the Zechstein and Scruff groups.

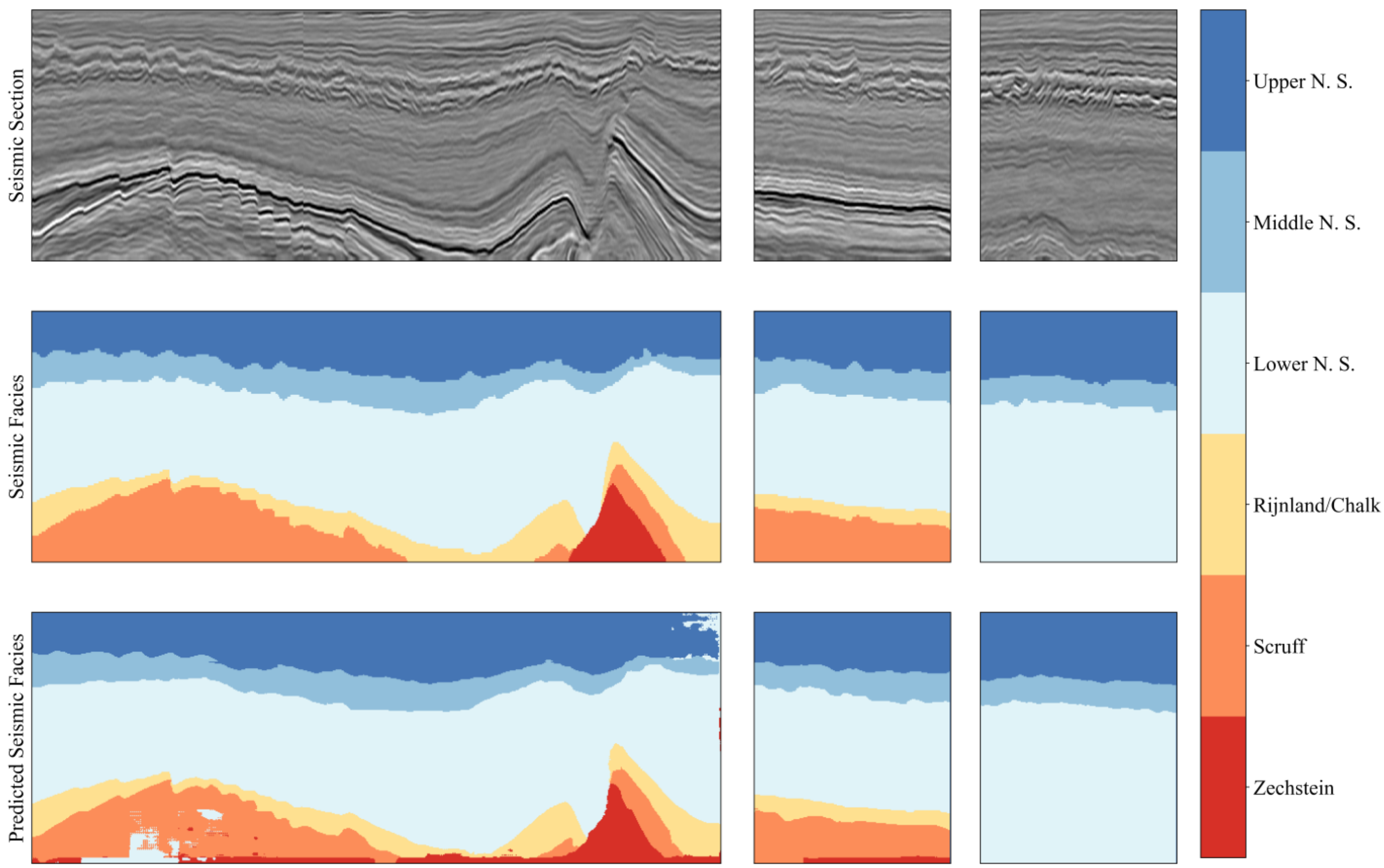

Fig. 7. Results obtained from EAN study. Top section: raw seismic section along inlines 295, crossline 620 from #Test1 and crossline 411 from #Test2 respectively. Middle section: Original labels and bottom section: Interpreted labels from *4 RB-TRB pairs with ASPP* model. We can see few misclassifications for inline 295 and crossline 620.

## E. *Model Training*

We performed several experiments with mini-batch sizes between 32 and 16. Tests were performed using the Adam and AdaDelta optimizers with different learning rates (i.e., $10^{-1}, 10^{-2}, 10^{-3}$, and $10^{-5}$) and a maximum number of epochs of 50. To address the problem of overfitting, we employed learning rate scheduling, dropout and early stopping. The learning rates decreased with the help of a learning rate scheduler when model performance did not increase for a few epochs. A dropout rate of 0.5 was used. Early stopping was used to stop the training process if performance of the model does not increase after a certain number of epochs. AdaDelta was used for EAN with ASPP module while Adam was used for other architectures. We used a weight decay parameter of 0.0001 for both Adam and AdaDelta optimizers while all other parameters were set to the default values. We trained the models with PyTorch 1.5.1 on Google Colaboratory using a 12GB NVIDIA Tesla K80 GPU.

## F. *Evaluation Metrics*

Several evaluation metrics for segmentation and classification derived from the confusion matrix were used to measure the performance of the proposed model.

*1) Segmentation:* To evaluate model performance, we used metrics including pixel accuracy (PA), class accuracy (CA), mean class accuracy (MCA), intersection over union (IoU), mean IoU (MIoU) and frequency weighted IoU (FWIoU), which are all commonly used metrics in the evaluation of computer vision models. $G_i$ represents the ground truth of pixels for class $i$, $P_i$ represents prediction for the class $i$ and $n_c$ represents the total number of the classes present. Metrics used in the present study are defined as follows:

a) **Pixel accuracy (PA)** is the percentage of total pixels correctly classified.

$$PA = \frac{\sum_i |P_i \cap G_i|}{\sum_i G_i} \tag{7}$$

b) **Class accuracy (CA)** is the percentage of total pixels correctly predicted for a particular class. We also used **mean class accuracy (MCA)** which is the average of CA.

$$CA = \frac{|P_i \cap G_i|}{G_i} \tag{8}$$

$$MCA = \frac{1}{n_c} \sum_i CA_i \tag{9}$$

c) **Intersection over union (IoU)**, also known as the **Jaccard Index**, is a measure of overlap between the predicted masks and the true masks. We also used **mean IoU (MIoU)**, calculated by averaging the IoU overall classes. IoU can be reformulated as the number of true positives (intersection) over the sum of true positives and false negatives. The metrics gives an estimate of how well the (x, y)-coordinates of our predicted bounding box is going to exactly match the (x, y)-coordinates of the ground-truth bounding box. Since the present dataset is of class imbalance, we weighed each class by its size, giving us the **frequency weighted IoU (FWIoU)**.

TABLE III
RESULTS OF EARTHADAPTNET WITH ITS VARIATION MODELS WHEN TESTED ON BOTH TEST SPLITS OF THE DATASET. ALL METRICS ARE IN THE RANGE (0-1), WITH LARGER VALUES BEING BETTER. ALL THE METRICS THAT HAVE OUTPERFORMED BASELINE MODEL [2] ARE IN BOLD NUMBERS. IN BASELINE MODEL, MIOU WASN'T REPORTED HENCE THE BOLD MIOU REPRESENTS BEST PERFORMING MODEL. 1 REPRESENTS UPPER NORTH SEA, 2 REPRESENTS MIDDLE NORTH SEA, 3 REPRESENTS LOWER NORTH SEA, 4 REPRESENTS RIJNLAND/CHALK, 5 REPRESENTS SCRUFF AND 6 REPRESENTS ZECHSTIEN

| Architecture | PA | MCA | FWIoU | MIoU | CA | | | | | |
|---|---|---|---|---|---|---|---|---|---|---|
| | | | | | 1 | 2 | 3 | 4 | 5 | 6 |
| Baseline [2] | 0.86 | 0.70 | 0.75 | - | 0.92 | 0.91 | 0.97 | 0.67 | 0.28 | 0.45 |
| DeepLab V3+ | 0.73 | 0.58 | 0.57 | 0.43 | 0.81 | 0.86 | 0.93 | 0.60 | 0.21 | 0.05 |
| 5 RB-TRB Pairs | 0.82 | **0.76** | 0.73 | 0.60 | 0.91 | 0.86 | 0.96 | **0.70** | 0.24 | **0.86** |
| 4 RB-TRB Pairs | 0.85 | 0.69 | 0.74 | 0.57 | **0.96** | 0.83 | 0.94 | **0.70** | **0.31** | 0.39 |
| 4 RB-TRB Pairs + ASPP | 0.85 | **0.78** | **0.77** | **0.62** | **0.95** | 0.84 | 0.94 | **0.73** | **0.35** | **0.84** |

$$IoU_i = \frac{|P_i \cap G_i|}{|P_i \cup G_i|} \tag{10}$$

$$MIoU = \frac{1}{n_c}\sum_i IoU_i \tag{11}$$

$$FWIoU = \frac{1}{\sum_i G_i} \cdot \sum_i |G_i| \cdot \frac{|P_i \cap G_i|}{|P_i \cup G_i|} \tag{12}$$

*2) Domain Adaptation:* The confusion matrix is used to calculate the number or frequency of true positive (TP), true negative (TN), false positive (FP), and false negative (FN). Results of DDA are evaluated using the following metrics:

a) **Class accuracy (CA)**measures how often the model predicts each class correctly. We also use **mean CA (MCA)** to save the checkpoints of our model. CA and MCA derived from the confusion matrix are calculated using the following equations:

$$CA = \frac{TP + TN}{TP + FP + TN + FN} \tag{13}$$

$$MCA = \frac{1}{n_c}\sum_i CA_i \tag{14}$$

b) **Precision** is a measure of the exactness of a model as it tells us that if a positive value is predicted by a model, how often is that prediction correct:

$$Precision = \frac{TP}{TP + FP} \tag{15}$$

c) **Recall,** also known as **true positive rate (TPR)** or **sensitivity**, is a measure of the completeness of positive outcomes of a model and it tells us that if the actual value is positive, how often the prediction is correct:

$$Recall = \frac{TP}{TP + FN} \tag{16}$$

d) **F1 Score** is the harmonic mean of precision and recall:

$$F1\,Score = \frac{Precision * Recall}{Precision + Recall} \tag{17}$$

e) **Pearson coefficient** ($r_p$) is well-known to researchers for the standard scenarioof normally distributed variables. $r_p$or bivariate correlation measures linear correlation between two variables X and Y forfinite sample sizes [36]. Correlation is a bivariate analysis that measures the strength of association between two variables and the direction of their relationship. In terms of the strength of relationship, the value of the correlation coefficient varies between +1 and -1. A value of ±1indicates a perfect degree of association between the two variables. As the correlation coefficient value goes towards 0, the relationship between the two variables will be weaker. The direction of the relationship is indicated by the sign of the coefficient; a + sign indicates a positive relationship and a – sign indicates a negative relationship.

TABLE IV
MEAN CLASS ACCURACY SHIFTS BETWEEN NETHERLANDS AND CANADA. N_CLASS → C_CLASS DEFINES THE SHIFT FROM THE FACIES CLASS FROM NETHERLANDS TO CANADA. "DIRECT TEST" REPRESENTS RESULTS FROM THE MODEL WITHOUT CORAL LOSS WHILE THE OTHER 3 MODELS ARE WITH CORAL LOSS.

| Experiment | | N_1 → C_1 | N_2 → C_2 | N_3 → C_3 |
|---|---|---|---|---|
| Direct Test | | 0.07 | 0.19 | 0.68 |
| DDA | 3RB+GAP+2FC | 0.12 | 0.61 | 0.81 |
| | 4RB+GAP+2FC | 0.11 | 0.58 | 0.84 |
| | 4RB+4FC | **0.19** | **0.75** | **0.91** |

f) $\boldsymbol{p-value}$ is used to estimate the linear relationship between two variables. In this study, a $p - value < 0.05$ refers to a statistically significant difference between variables and supports that two samples did not come from the same distribution. A $p - value > 0.05$ indicatesno statistically significant difference, and two samples come from the same distribution.

## V. EXPERIMENTAL RESULTS

### *A. Segmentation*

Our EAN model with ASPP comprises capabilities of U-Net, ResNet, DeepLab V3+ by incorporating the residual block and the ASPP module of UNet. For comparison with different classical segmentation architectures, we have added results from 4 types of models, i.e., simple UNet (Baseline), DeepLab V3+, modified UNet with residual and transposed residual blocks (5 RB-TRP Pairs and 4 RB-TRB Pairs) and finally, UNet with residual and transposed residual blocks as well as the ASPP module (EAN – 4 RB-TRB Pairs + ASPP) (Table III). EarthAdaptNet achieves an accuracy of >80% when applied to the 3 classes that belong to the North Sea Group, outperforming all other models compared (Table III). As shown in Table III, *4 RB-TRB Pairs with ASPP* and *5 RB-TRB Pairs* were able to give an overall accuracy of ~85% for the Zechstein class. This is a substantial improvement in comparison to performance of the baseline model used in one previous study [2].Although *4RB-TRB Pairs* was unable to capture the last 2 classes with comparable accuracy, with the introduction of an ASPP module, it has shown improved performance on those 2 classes as well. The results we present prove that the proposed model can capture the last 3 classes, which was not possible using the baseline model. In this study, one objective is to achieve improved results in minority facies classes (i.e., 4,5 and 6) of the dataset, and the EAN has shown higher accuracy for classes 4 and 6. The segmentation results are shown in Fig. 7. For the few pixels for the section inline 295 and crossline 620, misclassification occurred as the model was not able to efficiently identify scruffs for the North Sea groups.

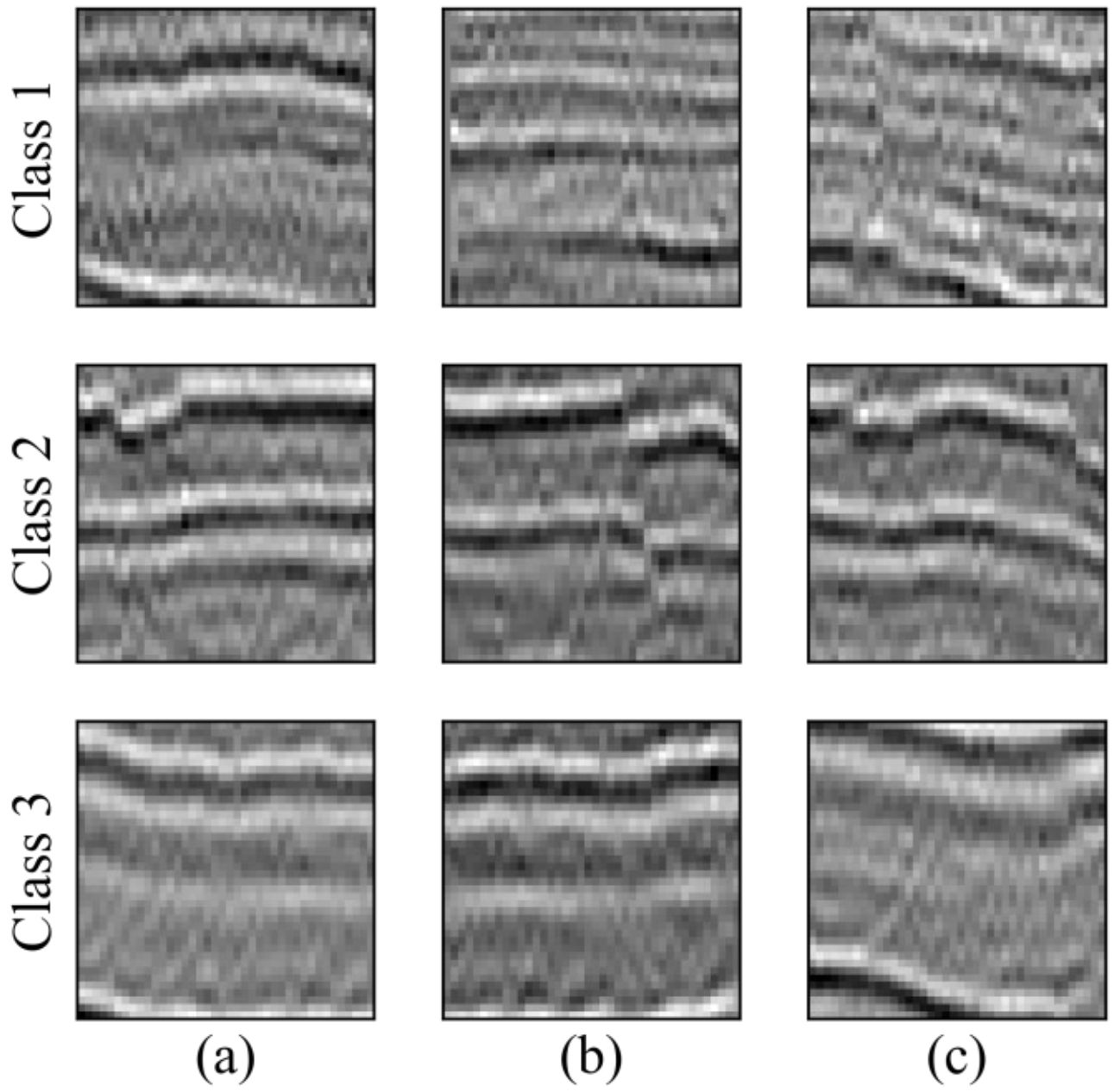

Fig. 8. A total of 3 test samples from each class. **(a)** Class 1: A patch along inline 1523 and crossline range [1320, 1360], with reflection at about 0.75 seconds. This sample was incorrectly predicted as Class 2. Class 2: A patch along inline 1523 and crossline range [1320, 1360], with reflection at about 0.9 seconds. This sample was correctly classified. Class 3: A patch along inline 1523 and crossline range [1320, 1360], with reflection at about 0.96 seconds. This sample was correctly classified. **(b)** Class 1: A patch along inline range [1555, 1595] and crossline 1121, with reflection at about 0.56 seconds. This sample was correctly classified. Class 2: A patch along inline range [1555, 1595] and crossline 1121, with reflection at about 0.9 seconds. This sample was correctly classified. Class 3: A patch along inline range [1555, 1595] and crossline 1121, with reflection at about 0.96 seconds. This sample was correctly classified. **(c)** Class 1: A patch along inline 1599 and crossline range [1450, 1490], with reflection at about 0.8 seconds. This sample was wrongly classified as Class 2. Class 2: A patch along inline 1599 and crossline range [1450, 1490], with reflection at about 0.94 seconds, which was wrongly classified as Class 3. Class 3: A patch along inline 1599 and crossline range [1450, 1490], with reflection at about 0.96 seconds, which was correctly predicted as Class 3.

TABLE V
STATISTICAL VARIANTS FOR *4RB+4FC* ARCHITECTURE. PEARSON COEFFICIENT AND $p-value$ DETERMINE HOW WELL DDA MODEL PERFORM.

| Entity | Pearson Coefficient | $p-value$ |
|---|---|---|
| *input_source – input_target* | 0.01 | 0.55 |
| *first_conv_source - first_conv _target* | 0.29 | 0.00 |
| *RB_1_source - RB_1_target* | 0.43 | 0.00 |
| *RB_2_source - RB_2_target* | 0.46 | 0.00 |
| *RB_3_source - RB_3_target* | 0.53 | 0.00 |
| *RB_4_source - RB_4_target* | 0.50 | 0.00 |
| *middle_conv_source - middle_conv_target* | 0.38 | 0.00 |
| *FC1_source - FC1_target* | 0.45 | 0.00 |
| *FC2_source - FC2_target* | 0.62 | 0.00 |
| *FC3_source - FC3_target* | 0.43 | 0.00 |
| *FC4_source – FC4_target* | 0.07 | 0.16 |

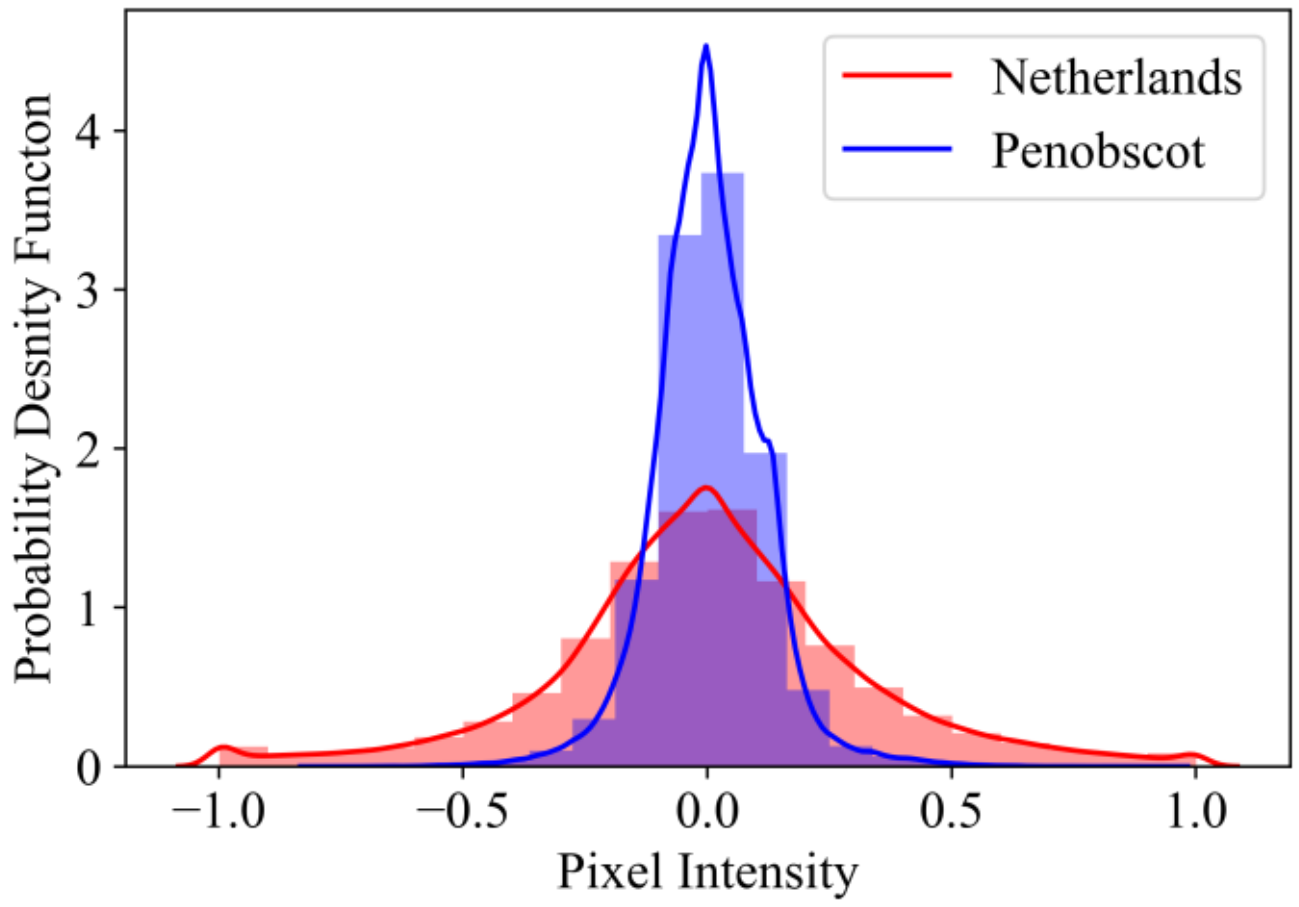

Fig. 9. Distribution in Netherlands and Canada for a batch size of 128. This batch is used to generate the feature map for the three variants of the DA model discussed. The initial distribution for Netherlands and Canada is significantly different. CORAL loss applies a non-linear transformation between TD and SD and the final classification result for TD has good accuracy.

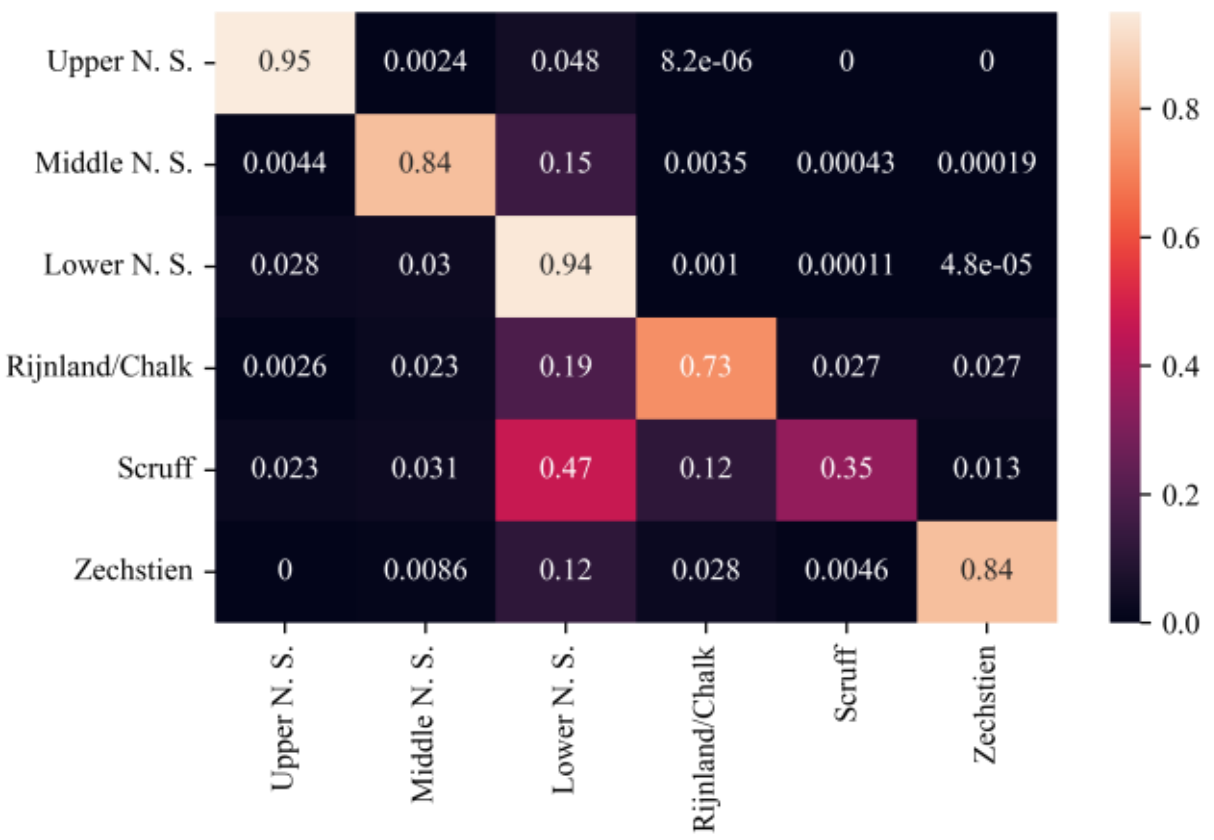

Fig. 10. Confusion Matrix derived from EAN-DDA network with 4 RB-TRB pairs + ASPP model.

A confusion matrix for the segmentation model, i.e., the EAN network, is shown in Fig. 10. From the confusion matrix it can be seen that Upper N. S. facies are sometimes misclassified as Middle and Lower N. S., while Middle N. S. is frequently misclassified as Lower N. S. In the meantime, Lower N. S. is sometimes misclassified as Upper., and Middle N. S. Rijnland/Chalk group is misclassified mostly as Lower N. S. group, and Scuff is most of the times misclassified as Lower N. S. and Rijnland/Chalk group. Lastly, Zechstien is sometimes classified as Lower N. S. group. Therefore, the confusion matrix not only tells us how our model performs, but also provides us information about how and where misclassification is happening, and this information could be used in future studies

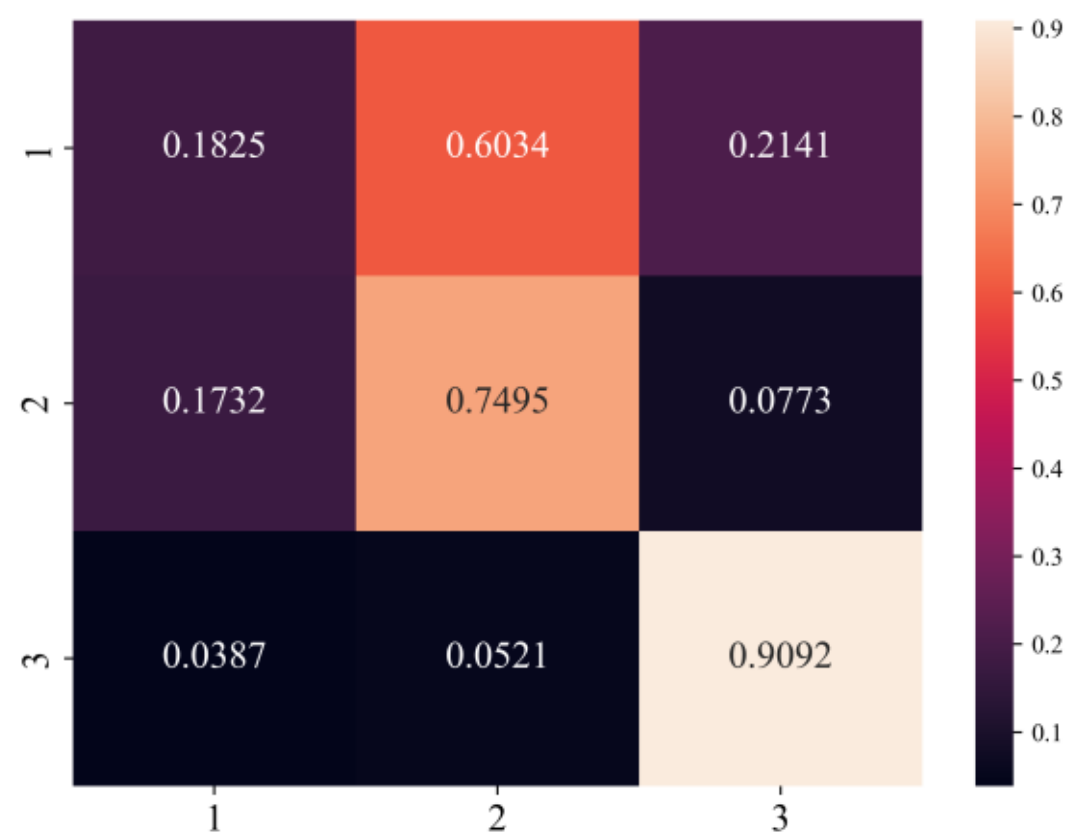

Fig. 11. Confusion Matrix derived from EAN network with 4 RB-TRB pairs + ASPP model. 4 RB + 4 FC layers.

for designing a better performing model.

### B. Deep Domain Adaptation

Experimental results obtained from the 3 DDA models are listed in Table IV, which demonstrate that the proposed unsupervised domain adaptation method effectively maps the feature space of TD (Canada) to that of SD (the Netherlands). To experimentally explore the potential upper bounds of the classification accuracy, we implemented three different model variants (Table I, Supplementary Table I, and Supplementary Table II). To address the severe domain shift problem in seismic image analysis, we applied a classifier trained on the Netherlands domain to the Canada domain through a *direct test*, which was defined as a scenario where the weights of a model trained on the Netherlands F3 data are directly used to predict labels for the Penobscot dataset (note that we do not use CORAL Loss in the *direct test*). The model that achieves the best performance on the direct test completely failed on Canada data, with a maximum accuracy of 0.07, 0.19, and 0.68 respectively for the three classes. This leads to the conclusion that although reflection patterns from the Netherlands and Canada are similar, i.e., sharing high level representations and identical label space, the significant difference in their low-level characteristics makes it extremely difficult for models trained on the Netherlands data to effectively extract features from the Canada data.

With our unsupervised DDA method, we observe a great improvement in the classification performance on the TD (Canada) compared with that achieved by the direct test on the TD (Canada). We believe that the DDA approach will be able to address the problems of data scarcity, as well as domain shift between classes from 2 different domains. We studied three different model variations of the EAN-DDA model as discussed in section II-A. A detailed comparison between all three models has been presented in discussion sections as well. The architecture 4RB+4FC gave better performance compared to other model variants and results collected from this model are discussed. The EAN-DDA model has increased the average accuracy to ~30% to 40% for facies classes in Canada. Class H4-H3 (C_3) of Canada can be best correlated with the Zechstien group (N_3) of the Netherlands and is detected with an accuracy of 91%, demonstrating an increase in accuracyby23% compared to the direct test. Class H5-H4 (C_2) of Canada can be correlated with the Scuff group (N_2) of the Netherlands and is detected with an accuracy of 75%, indicating an increase in accuracy by56% compared to the direct test. Class H5-H6 (C_1) of Canada can be correlated with the Chalk and Rijnland groups (N_1) of the Netherlands and is detected with an accuracy of 19%, showing an increase in accuracy by12% compared to the direct test (Table IV). The reflection patterns of Chalk and Rijnland (N_1) and H5-H6 (C_1) does not match properly (Table II) as discussed in the dataset section, making the EAN-DDA model unable to achieve good performance for class 1. Fig. 8 shows an example of 3 patches for each class used in the testing phase and their respective class. Given that the goal of the DDA method is to map SD and TD to a domain invariant feature space, we have successfully matched the distributions of SD and TD across the networks (Fig. 13, Supplementary Fig. 1, and Supplementary Fig. 2). The $r_p$ and $p - value$ between feature maps of SD and TD are reported in Table V, Supplementary Table III, and Supplementary Table IV.

In Fig. 11, we presented a confusion matrix of the classification model, i.e., the EAN-DDA network. From the confusion matrix, it can be observed that representative class 1 is commonly misclassified as representative class 2, and sometimes as representative class 3, while representative class 2 is mostly misclassified as representative 1. A representative class 3 shows best accuracy and rarely gets misclassified as other representative classes. Low accuracy in the classification of class 1 is attributed to the fact that reflection patterns don't match properly and have some difference. This point can be taken into consideration in future studies.

## VI. Discussion

### A. Optimizer Selection

To perform a comparison between the models tested, a mini-batch size of 32 is used. We store the model weights for best performing epoch and the definition of best performing model was based on the best MCA achieved by the model. Experiments are performed primarily to analyze the behavior of different optimizers including Adam and AdaDelta, and as mentioned in section 4D, to fine tune the decay parameter. AdaDelta and Adam are very similar algorithms that perform comparably well in similar circumstances. However, based on [37], we infer that the bias-correction in the Adam optimizer slightly outperforms towards the end of the optimization process as gradients become sparser. Hence, Adam has been suggested to be a better overall choice over AdaDelta [38]. The second set of experiments was performed to determine an epoch size, so the model training does not reach timeout. The models were tested for 50 and 100 epochs respectively, where we did not observe a significant increase in CA (~2%) of the North Sea group.

### B. Patch-based models

The model architectures used in this study are patch-based, i.e., trained on patches of different depths, since the spatial dependencies are lost the architectures which led to confusion between facies classes. Section-based models are superior to patch-based models given their ability to incorporate spatial and contextual information within each seismic section. However, because of the unknown and large size of the sections, computations can become very slow. Due to lack of computational resources and due to that some artifacts were introduced such as Scuff group being misclassified, we used a patch-based model.

In the EAN study, using ASPP block as a bridging layer, and data augmentation, we overcome the problem of class imbalance and the accuracy for classifying the minority classes, such as the Zechstein, Scruff groups and Rijnland/Chalk group, improved by 39%, 7%and 6%, respectively [2].

In comparison to the patch-based baseline model [2], EAN converges faster when the same dataloader is used. For classes 1, 4, 5 and 6, the EAN outperformed the baseline model by 4%, 6%, 7%, and 39%, respectively, as measured by accuracy.

### C. *Encoder - Decoder Architecture*

EarthAdaptNet uses RBs [25] to extract features (encoder) and transposed units of the same structure to reconstruct the original image with its respective labels (decoder). The skip connections skip some layers in the neural network and feed the output of one given layer as the input to the next few layers, instead of only the next immediate layer. It provides an alternative path for the gradient (with backpropagation) to flow through the network, and these additional paths are beneficial for the convergence of models during training. The skip connection incorporates the topology to combine coarse information with detailed, lower layer information. In the present study, the 4 RBs architecture has a downsampled image size of $7X7$ whereas the 5 RBs architecture has a downsampled image size of $3X3$. Hence the 4 RBs architecture has access to more samples, which eventually leads to a higher MCA.

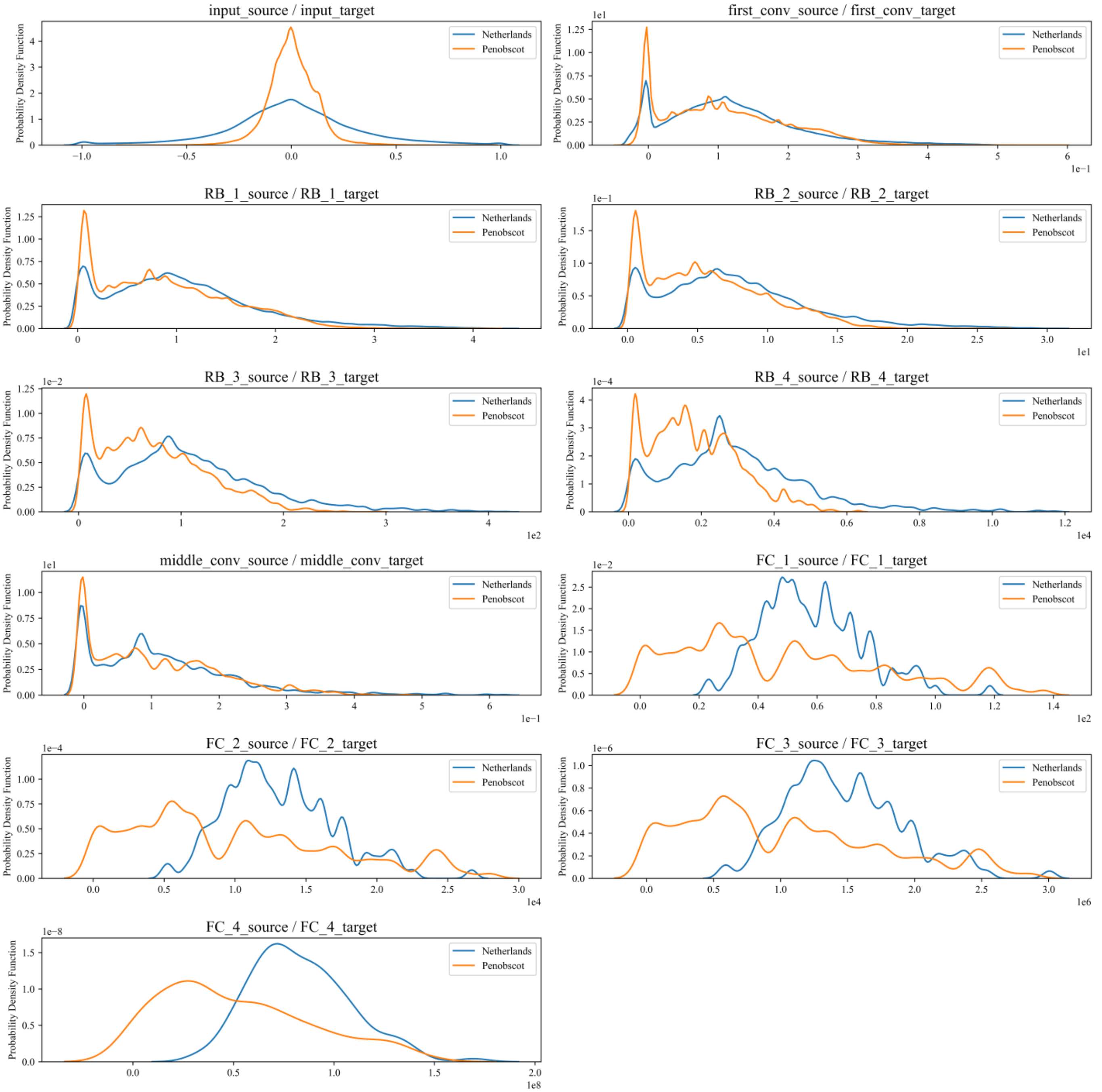

Fig. 12. Comparison of the feature maps distributions from source and target domains when no CORAL Loss is applied, i.e., with random initialisation. Distributions from the layer for the architecture with 4 Encoders followed by 4 FC layers (*4RB+4FC)* shows how the SD (Netherlands) and TD (Canada) distributions do not correlate when the model is not trained.

The 4 RBs with ASPP achieve a better performance with an MCA of 78%, and a Mean IoU of 62% which shows that the spatial coordinates are predicted slightly more accurately and the ASPP greatly enlarges the Valid Receptive Field (VRF). However, due to the cropping size, the VRF field is also limited to local patches, where not enough context information is included. The downscaled images used in the first training stage greatly enlarge the VRF. VRF is known to be crucial for the image-based classification of objects [39], and if the VRF is large enough to represent the whole object and its surroundings, a DNN architecture learns the potential representations of their correlations more efficiently [39].The proposed architecture was able to give a MCA of ~85% for the Zechstein class. Our EAN model has a smaller number of parameters as well when compared to the baseline model. The baseline model has 84,031,366 parameters while our best performing model, i.e., 4 RB-TRB Pairs + ASPP has 11,940,317 parameters which is about an 8-fold decrease in the number of parameters and hence the size of the model. The baseline model converged after 16 hours of training. Our model took about 6 hours for convergence even though we were using a lower performing GPU (Nvidia Tesla K80) as compared to baseline (Nvidia Titan X).

### *D. Distribution of feature maps*

To study the performance of the EAN-DDA architecture and how its various components affect the accuracy of the classification of each class, we plot the probability distribution of feature maps. A probability distribution is a statistical function that describes all the possible outcome values and likelihoods that a random variable can take within a given range. In statistical studies, the null hypothesis is a default hypothesis that a quantity (typically the difference between two situations) to be measured is zero (null). In this scenario, the null hypothesis is to determine if there is an indication that the samples from TD and SD are derived from different distributions. The $p-value$ is defined as the probability of obtaining test results at least as extreme as the results observed, assuming that the null hypothesis is correct.

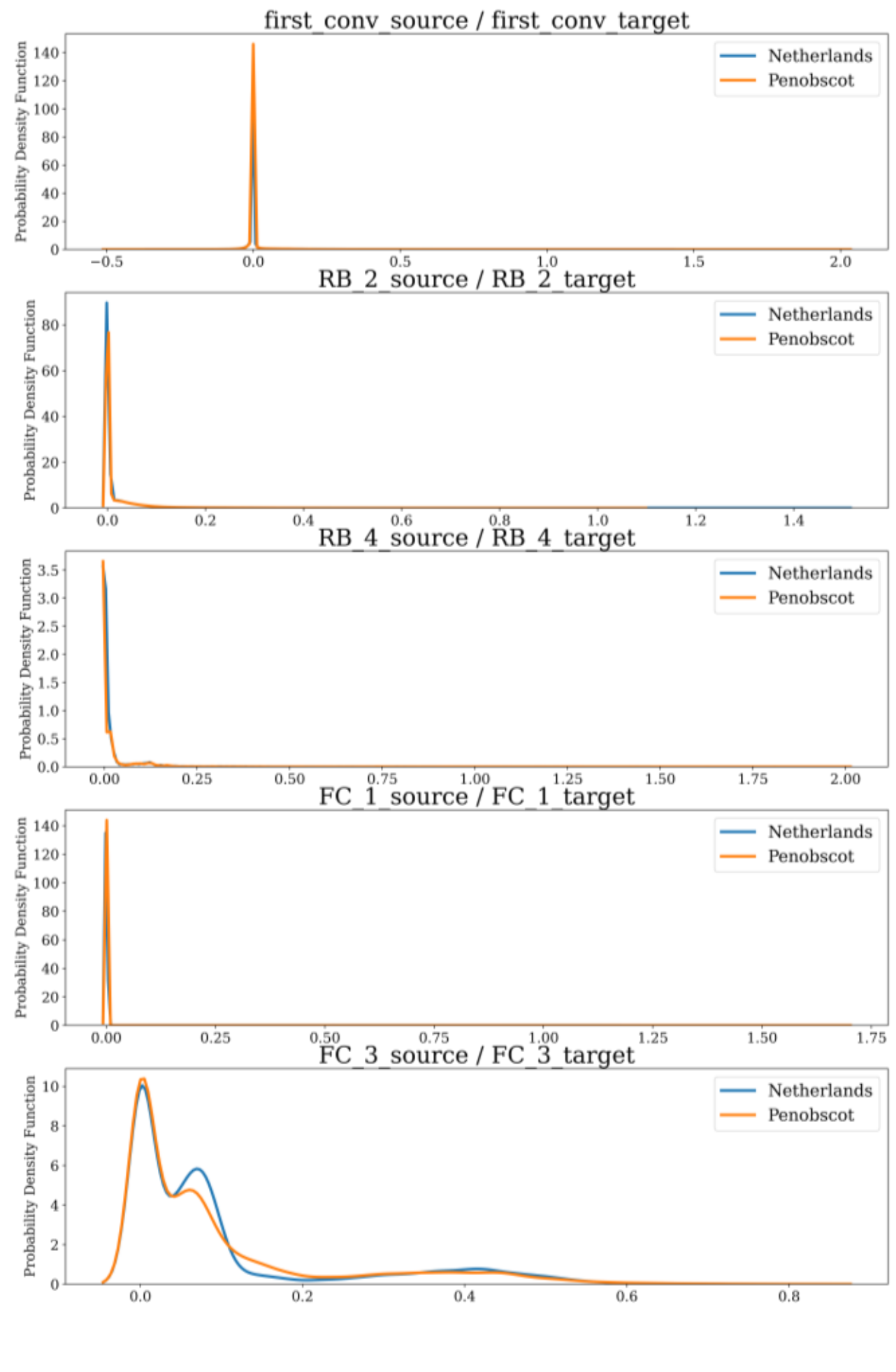

Fig. 13. Comparison of the feature maps distributions from source and target domains. Distributions from the layer for the architecture with 4 Encoders followed by 4 FC layers (*4RB+4FC)* shows how the SD (Netherlands) and TD (Canada) distributions correlate.

A $p-value < 0.05$ is sufficient to reject the null hypothesis and conclude that a significant difference between the two distributions does exist. To measure the similarity among feature maps, a batch size of 128 was randomly selected and feature maps were generated for all the three models viz:*4RB+4FC*, *4RB+GAP+2FC* and *3RB+GAP+2FC*. The similarity of the feature map distributions is calculated using the Pearson coefficient $r_p$ and $p-value$.

Here, we present the feature map distribution using a two-stage process: First, we show the feature map of an untrained EAN-DDA network, i.e., without CORAL loss (Fig. 12), and we then show the feature map after EAN-DDA network is trained, to show how much contribution is introduced by CORAL Loss (Fig. 13). Fig. 12 shows the distribution of feature maps of the *4RB+4FC* architecture before training the EAN-DDA network. As can be seen, distribution of SD and TD differs across networks.

Fig. 13shows the distribution of feature maps of the *4RB+4FC*architecture after training for 100 epochs. The similarity improves in deeper layers as indicated by the $r_p$ values. Furthermore, a $p-value < 0.05$ shows that the similarity of distributions is statistically significant (Table V). The $r_p$ value of all layers except input and last FC layer is high, indicating the features from source and target correlate. We can clearly see that adding the CORAL loss helps achieve better performance on the target domain while maintaining strong classification accuracy on the source domain [23].

Supplementary Fig. 1 shows the distribution of output feature maps for the*4RB+GAP+2FC*architecture after training for 100 epochs. Based on Supplementary Table III, the RBs output feature map correlation increases as we go deeper into the model. The feature maps of FC layers at the start are the same, making this model not very efficient at capturing domain invariant features as compared to the *4RB+4FC* model. However, this model has a higher $r_p$ value than the *3RB+GAP+2FC*architecture. The $p-value$ for all the distributionsexcept input and last layer is $< 0.05$, hence the source-target distributions are statistically significant therefore rejecting the null hypothesis. Supplementary Fig. 2 shows the distribution of output feature maps for the *3RB+GAP+2FC*architecture after training for 100 epochs. Based on Supplementary Table IV, the output feature maps are not well correlated, indicated by low$r_p$ values. The highest $r_p$ value is observed from *GAP_source – GAP_target* as feature maps in deeper layers start getting more similar. The $p-value$ for all the distribution except input and last layer is $< 0.05$, hence the source-target distributions are statistically significant.

### E. Effect of Global Average Pooling

GAP is an operation that calculates the average output of feature maps in the previous layer. This operation reduces the data load significantly and prepares the model for the final classification layer. Since GAP does not contain anything trainable, the $r_p$ between the source and target might not be well correlated as shown in models *4RB+GAP+2FC* and *3RB+GAP+2FC*.

## VII. Conclusion

In conclusion, we have introduced a deep learning model named EarthAdaptNet (EAN) that outperformed segmentation models such as UNet, UNet + ResNet and DeepLab V3+. It can efficiently classify facies with patch sections and is able to achieve a classification accuracy>50% for smaller classes like Zechstein and Scruff. We demonstrate that the EAN was able to achieve higher accuracy in the classification of minority classes in comparison with baseline models. The architecture performs better than the patch-based baseline model. We present and describe an effective and efficient method for unsupervised domain adaptation using the CORrelation ALignment (CORAL) method. The CORAL method minimizes domain shift by aligning the second-order statistics of source and target distributions, without requiring any target labels. The CORAL loss applied to domain adaptation algorithms is then extended to EarthAdaptNet. The proposed DDA approach is among one of the first applications of DDA to the study of unlabeled seismic facies. As the result, two out of the three classes of Penobscot were classified with an accuracy >75%. We also present and examine 3 variants of the proposed DDA architecture to understand how the components such as residual blocks, global average pooling, and fully connected layers behave in domain adaptation. We note that the 4RB+4FCDDA model shows promising results in class 2 and class 3. We observe that more complex classifier modules (i.e., classifier modules with a greater number of FC layers) lead to higher accuracy. The MCA of 4RB+4FC is ~62% while the MCA of 4RB+GAP+2FC is ~51%. In the meantime, accuracy decreased following a decrease in the number of RBs used in the contracting path: the MCA of 3RB+GAP+2FCis ~50%, which is the lowest among all the three EAN-DDA model variants. Domain transfer for class 1 of SD to TD is not efficient since the reflection patterns of the Chalk and Rijnland group of SD do not match with that of theH5-H6 of TD. The average MCA of all presented EAN-DDA models is ~54%. The feature map distribution study of SD and TD proved useful to verify how well the target domain is adapting. We show that DDA has the potential to achieve high performance when labelled data are scarce, or when subject matter experts are not available for the generation of labelled data.

One limitation of patch-based models for segmentation is that these models don't get to see the whole seismic section at a time. Instead, it only looks at a patch and loses spatial information. To overcome these difficulties in the patch-based model we propose (1) metadata tagging and (2) the use of architectures like Recurrent Neural Networks, to preserve spatial information in future studies, as seismic data is essentially a time series data and it already incorporates spatial information.

The present study applies DDA methods to the seismic reflection patterns. The study can be potentially extended to validate and understand the generalizability of the proposed approach to different geological domains. Deep Domain Adaptation can be used to study other seismic attributes such as Direct Hydrocarbon Indicators (DHIs) like bright spot from seismic data and hydrocarbon detection from well logs. Future research directions will include (1) conversion of classification problem to segmentation problem (2) EAN-DDA study is a Divergence-based DDA methodology which specifically is designed for classification. For segmentation, one can use Adversarial-based DDA such as CoGAN [40] and Pixel-level Domain Transfer [41].

## ACKNOWLEDGMENT

This research was supported and funded by LiveAI. The facilities of F3 seismic Data Services were accessed from olivesgatech GitHub handle and for Penobscot data from zenodo.org by Baroni 2018. Data for F3 and Penobscot are freely available for research purpose. We thank Md Rifat Arefin for providing comments and suggestions on model training.

## REFERENCES

[1] Alaudah, Y., S. Gao, and G. AlRegib, *Learning to label seismic structures with deconvolution networks and weak labels*, in *SEG Technical Program Expanded Abstracts 2018*. 2018, Society of Exploration Geophysicists. p. 2121-2125.

[2] Alaudah, Y., et al., *A machine-learning benchmark for facies classification.* Interpretation, 2019. **7**(3): p. SE175-SE187.

[3] Di, H., Z. Wang, and G. AlRegib, *Real-time seismic-image interpretation via deconvolutional neural network*, in *SEG Technical Program Expanded Abstracts 2018*. 2018, Society of Exploration Geophysicists. p. 2051-2055.

[4] Dramsch, J.S. and M. Lüthje, *Deep-learning seismic facies on state-of-the-art CNN architectures*, in *SEG Technical Program Expanded Abstracts 2018*. 2018, Society of Exploration Geophysicists. p. 2036-2040.

[5] Huang, L., X. Dong, and T.E. Clee, *A scalable deep learning platform for identifying geologic features from seismic attributes.* The Leading Edge, 2017. **36**(3): p. 249-256.

[6] Shi, Y., X. Wu, and S. Fomel, *Automatic salt-body classification using a deep convolutional neural network*, in *SEG Technical Program Expanded Abstracts 2018*. 2018, Society of Exploration Geophysicists. p. 1971-1975.

[7] Waldeland, A.U. and A. Solberg. *Salt classification using deep learning*. European Association of Geoscientists & Engineers.

[8] Zhao, T., *Seismic facies classification using different deep convolutional neural networks*, in *SEG Technical Program Expanded Abstracts 2018*. 2018, Society of Exploration Geophysicists. p. 2046-2050.

[9] Araya-Polo, M., et al., *Automated fault detection without seismic processing.* The Leading Edge, 2017. **36**(3): p. 208-214.

[10] Alaudah, Y. and G. AlRegib. *Weakly-supervised labeling of seismic volumes using reference exemplars*. IEEE.

[11] Coléou, T., M. Poupon, and K. Azbel, *Unsupervised seismic facies classification: A review and comparison of techniques and implementation.* The Leading Edge, 2003. **22**(10): p. 942-953.

[12] de Matos, M.C., P.L. Osorio, and P.R. Johann, *Unsupervised seismic facies analysis using wavelet transform and self-organizing maps.* Geophysics, 2007. **72**(1): p. P9-P21.

[13] Dubois, M.K., G.C. Bohling, and S. Chakrabarti, *Comparison of four approaches to a rock facies classification problem.* Computers & Geosciences, 2007. **33**(5): p. 599-617.

[14] Civitarese, D., et al., *Semantic segmentation of seismic images.* arXiv preprint arXiv:1905.04307, 2019.

[15] Hagos, M.T. and S. Kant, *Transfer learning based detection of diabetic retinopathy from small dataset.* arXiv preprint arXiv:1905.07203, 2019.

[16] Chevitarese, D., et al., *Transfer learning applied to seismic images classification.* AAPG Annual and Exhibition, 2018.

[17] Dou, Q., et al., *Unsupervised cross-modality domain adaptation of convnets for biomedical image segmentations with adversarial loss.* arXiv preprint arXiv:1804.10916, 2018.

[18] Yosinski, J., et al. *How transferable are features in deep neural networks?*

[19] Zamir, A.R., et al. *Taskonomy: Disentangling task transfer learning*.

[20] Gretton, A., et al., *Covariate shift by kernel mean matching.* Dataset shift in machine learning, 2009. **3**(4): p. 5.

[21] Torralba, A. and A.A. Efros. *Unbiased look at dataset bias*. IEEE.

[22] Patel, V.M., et al., *Visual domain adaptation: A survey of recent advances.* IEEE signal processing magazine, 2015. **32**(3): p. 53-69.

[23] Sun, B. and K. Saenko. *Deep coral: Correlation alignment for deep domain adaptation*. 2016. Springer.

[24] Ronneberger, O., P. Fischer, and T. Brox. *U-net: Convolutional networks for biomedical image segmentation*. Springer.

[25] He, K., et al. *Deep residual learning for image recognition*.

[26] Wojna, Z., et al., *The devil is in the decoder: Classification, regression and gans.* International Journal of Computer Vision, 2019. **127**(11-12): p. 1694-1706.

[27] Chen, L.-C., et al., *Rethinking atrous convolution for semantic image segmentation.* arXiv preprint arXiv:1706.05587, 2017.

[28] Zhao, Q., et al., *Interpretable Relative Squeezing bottleneck design for compact convolutional neural networks model.* Image and Vision Computing, 2019. **89**: p. 276-288.

[29] Mattos, A.B., et al. *Assessing texture descriptors for seismic image retrieval*. IEEE.

[30] Van Wagoner, J.C., et al., *Seismic stratigraphy interpretation using sequence stratigraphy: Part 2: key definitions of sequence stratigraphy.* 1987.

[31] Baroni, L.a.S., Reinaldo Mozart and Ferreira, Rodrigo and Chevitarese, Daniel and Szwarcman, Daniela and Brazil, Emilio Vital, *Penobscot Interpretation Dataset*. 2018.

[32] Vail, P.R., R.M. Mitchum Jr, and S. Thompson Iii, *Seismic stratigraphy and global changes of sea level: Part 4. Global cycles of relative changes of sea level.: Section 2. Application of seismic reflection configuration to stratigraphic interpretation.* 1977.

[33] Posamentier, H.W. and P.R. Vail, *Eustatic controls on clastic deposition II—sequence and systems tract models.* 1988.

[34] Reading, H.G. and M. Richards, *Turbidite systems in deep-water basin margins classified by grain size and feeder system.* AAPG bulletin, 1994. **78**(5): p. 792-822.

[35] Knebel, H.J., *Holocene depositional history of a large glaciated estuary, Penobscot Bay, Maine.* Marine Geology, 1986. **73**(3-4): p. 215-236.

[36] de Winter, J.C.F., S.D. Gosling, and J. Potter, *Comparing the Pearson and Spearman correlation coefficients across distributions and sample sizes: A tutorial using simulations and empirical data.* Psychological methods, 2016. **21**(3): p. 273.

[37] Kingma, D.P. and J.L. Ba. *Adam: A method for stochastic gradient descent*.

[38] Ruder, S., *An overview of gradient descent optimization algorithms.* arXiv preprint arXiv:1609.04747, 2016.

[39] Ding, L., J. Zhang, and L. Bruzzone, *Semantic Segmentation of Large-Size VHR Remote Sensing Images Using a Two-Stage Multiscale Training Architecture.* IEEE Transactions on Geoscience and Remote Sensing, 2020.

[40] Liu, M.-Y. and O. Tuzel. *Coupled generative adversarial networks*. 2016.

[41] Yoo, D., et al. *Pixel-level domain transfer*. 2016. Springer.

[42] Hannun, Awni, Chuan Guo, and Laurens van der Maaten. "Measuring Data Leakage in Machine-Learning Models with Fisher Information." *arXiv preprint arXiv:2102.11673* (2021).

[43] Zheng, Alice, and Amanda Casari. *Feature engineering for machine learning: principles and techniques for data scientists*. " O'Reilly Media, Inc.", 2018

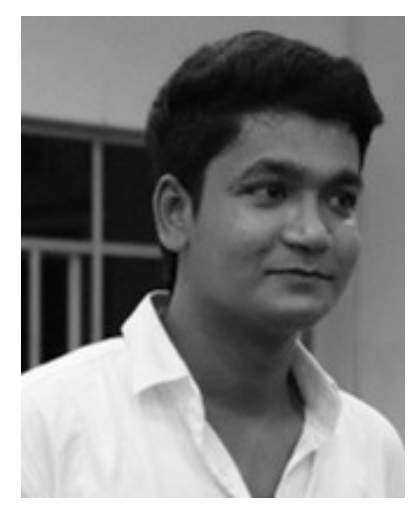

**M Quamer Nasim** is a Ph.D. candidate at the Indian Institute of Technology, Kharagpur, India, working on applications of AI in hydrocarbon exploration. He completed his Master's Degree in Geophysics and Bachelor's Degree in Geology. He is currently working as a Sr AI Researcher at DeepKapha.ai. His work and interests revolve around Computer Vision and his expertise lies in integrating Earth Sciences and AI.

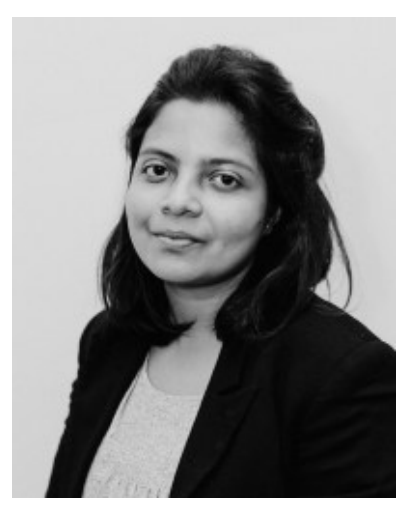

**Tannistha Maiti** received her Ph.D. degree in Geophysics from the University of Calgary, Canada in 2018. She is presently an AI researcher at DeepKapha.ai. Her research concentrates on deep learning in various earth science domains.

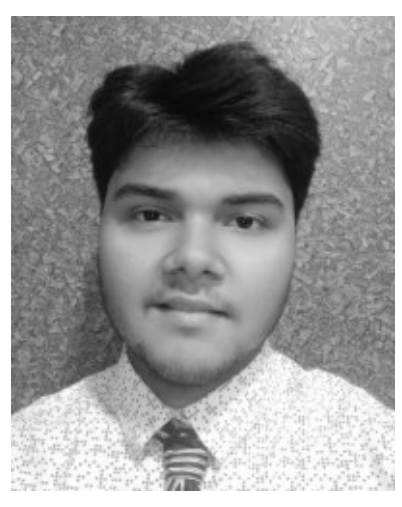

**Ayush Srivastava** is a workaholic individual currently pursuing a Bachelor's degree focused in Mechanical Engineering from Indian Institute of Technology, Kharagpur with a Minor in Computer Science and Engineering. A working professional with a history of working on a wide spectrum of domains spanning Artificial Intelligence and Software Development, He brings to the table a working experience of more than 3 years in the industry as well as academia.

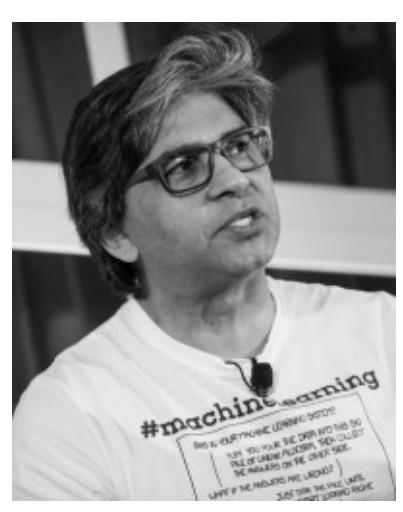

**Tarry Singh** is an experienced AI entrepreneur, researcher, educator and data expert with over 20 years' experience in multiple industries. He is a subject matter expert with industry knowledge in deep learning, machine learning and artificial intelligence. He is an internationally acknowledged speaker, author, mentor and startup coach. Tarry has (co-)authored several publications and books – in research (IEEE, Springer, ACM etc.) as well as practical industry areas. He is passionate about using technology for social good and works with various organizations such as World Health Organization, NGOs, United Nations towards humanitarian goals.

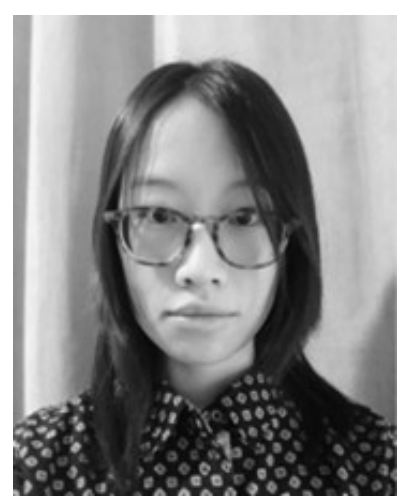

**Jie Mei** received her M.S. degree in cognitive neuroscience and computational neuroscience from École Normale Supérieure Paris, France in 2015, and completed her Ph.D. degree in medical neuroscience at Charité – Universitätsmediz in Berlin, Germany, in 2019. She is currently a postdoctoral fellow at University of Quebec. Her major research interests include computational neuroscience, deep learning, and applications of deep learning techniques for image classification.

# Supplementary Material

SUPPLEMENTARY TABLE I

LAYERWISE ARCHITECTURAL DESCRIPTION OF THE FEATURE MAPS FOR THE SOURCE AND TARGET DOMAINS FOR THE MODEL WITH 4 ENCODERS FOLLOWED BY A GAP LAYER AND 2 FC LAYERS (*4RB+GAP+2FC*).

| Entity | Entity Size | Entity Description |
|---|---|---|
| *input_source* | (128, 1, 40, 40) | SD image batch |
| *input_target* | (128, 1, 40, 40) | TD image batch |
| *first_conv_source* | (128, 64, 40, 40) | SD image batch feature map after first conv layer |
| *first_conv _target* | (128, 64, 40, 40) | TD image batch feature map after first conv layer |
| *RB_1_source* | (128, 64, 20, 20) | SD image batch feature map after first RB |
| *RB_1_target* | (128, 64, 20, 20) | TD image batch feature map after first RB |
| *RB_2_source* | (128, 128, 10, 10) | SD image batch feature map after second RB |
| *RB_2_target* | (128, 128, 10, 10) | TD image batch feature map after second RB |
| *RB_3_source* | (128, 256, 5, 5) | SD image batch feature map after third RB |
| *RB_3_target* | (128, 256, 5, 5) | TD image batch feature map after third RB |
| *RB_4_source* | (128, 512, 3, 3) | SD image batch feature map after fourth RB |
| *RB_4_target* | (128, 512, 3, 3) | TD image batch feature map after fourth RB |
| *GAP_source* | (128, 512, 1, 1) | SD image batch feature map after GAP layer |
| *GAP_target* | (128, 512, 1, 1) | TD image batch feature map after GAP layer |
| *FC_1_source* | (128, 256) | SD image batch feature map after first FC layer |
| *FC_1_target* | (128, 256) | TD image batch feature map after first FC layer |
| *FC_2_source* | (128, 3) | SD image batch feature map after last FC layer |
| *FC_2_target* | (128, 3) | TD image batch feature map after last FC layer |

SUPPLEMENTARY TABLE II

LAYERWISE ARCHITECTURAL DESCRIPTION OF THE FEATURE MAPS FOR THE SOURCE AND TARGET DOMAINS FOR THE MODEL WITH 3 ENCODERS FOLLOWED BY A GAP LAYER AND 2 FC LAYERS (*3RB+GAP+2FC*).

| Entity | Entity Size | Entity Description |
|---|---|---|
| *input_source* | (128, 1, 40, 40) | SD image batch |
| *input_target* | (128, 1, 40, 40) | TD image batch |
| *first_conv_source* | (128, 64, 40, 40) | SD image batch feature map after first conv layer |
| *first_conv _target* | (128, 64, 40, 40) | TD image batch feature map after first conv layer |
| *RB_1_source* | (128, 64, 20, 20) | SD image batch feature map after first RB |
| *RB_1_target* | (128, 64, 20, 20) | TD image batch feature map after first RB |
| *RB_2_source* | (128, 128, 10, 10) | SD image batch feature map after second RB |
| *RB_2_target* | (128, 128, 10, 10) | TD image batch feature map after second RB |
| *RB_3_source* | (128, 256, 5, 5) | SD image batch feature map after third RB |
| *RB_3_target* | (128, 256, 5, 5) | TD image batch feature map after third RB |
| *GAP_source* | (128, 512, 1, 1) | SD image batch feature map after GAP layer |
| *GAP_target* | (128, 512, 1, 1) | TD image batch feature map after GAP layer |
| *FC_1_source* | (128, 128) | SD image batch feature map after first FC layer |
| *FC_1_target* | (128, 128) | TD image batch feature map after first FC layer |
| *FC_2_source* | (128, 3) | SD image batch feature map after last FC layer |
| *FC_2_target* | (128, 3) | TD image batch feature map after last FC layer |

SUPPLEMENTARY TABLE III

STATISTICAL RESULTS OF THE*4RB+GAP+2FC* ARCHITECTURE. PEARSON CORRELATION COEFFICIENT AND P-VALUE DETERMINING HOW WELL THE DDA MODEL PERFORMS ARE SHOWN.

| Entity | Pearson Correlation Coefficient | p-value |
|---|---|---|
| *input_source - input_target* | 0.01 | 0.55 |
| *first_conv_source - first_conv _target* | 0.17 | 0.00 |
| *RB_1_source - RB_1_target* | 0.31 | 0.00 |
| *RB_2_source - RB_2_target* | 0.42 | 0.00 |
| *RB_3_source - RB_3_target* | 0.61 | 0.00 |
| *RB_4_source - RB_4_target* | 0.34 | 0.00 |
| *GAP_source - GAP_target* | 0.34 | 0.00 |
| *FC_1_source – FC_1_target* | 0.15 | 0.00 |
| *FC_2_source – FC_2_target* | 0.15 | 0.01 |

SUPPLEMENTARY TABLE IV

STATISTICAL RESULTS OF THE *3RB+GAP+2FC* ARCHITECTURE. PEARSON CORRELATION COEFFICIENT AND P-VALUE DETERMINING HOW WELL THE DDA MODEL PERFORMS ARE SHOWN.

| Entity | Pearson Correlation Coefficient | p-value |
|---|---|---|
| *input_source - input_target* | 0.01 | 0.55 |
| *first_conv_source - first_conv _target* | 0.02 | 0.00 |
| *RB_1_source - RB_1_target* | 0.11 | 0.00 |
| *RB_2_source - RB_2_target* | 0.18 | 0.00 |
| *RB_3_source - RB_3_target* | 0.37 | 0.00 |
| *GAP_source - GAP_target* | 0.42 | 0.00 |
| *FC_1_source – FC_1_target* | 0.20 | 0.00 |
| *FC_2_source – FC_2_target* | 0.18 | 0.01 |

SUPPLEMENTARY TABLE V

RESULTS OF THE EARTHADAPTNET WITH ITS VARIATION HAVING THE BATCH NORMALIZATION LAYER IN SKIP CONNECTION, OBTAINED FROM THE TEST SPLIT OF THE DATASET. ALL METRICS ARE WITHIN THE RANGE (0-1), WITH LARGER VALUES REPRESENTING BETTER RESULTS.

| Architecture | PA | MCA | FWIoU | MIoU | CA | | | | | |
|---|---|---|---|---|---|---|---|---|---|---|
| | | | | | 1 | 2 | 3 | 4 | 5 | 6 |
| 4 RB-TRB Pairs | 0.05 | 0.24 | 0.02 | 0.03 | 0.00 | 0.00 | 0.01 | 0.38 | 0.04 | 0.99 |
| 4 RB-TRB Pairs + ASPP | 0.09 | 0.28 | 0.03 | 0.04 | 0.01 | 0.07 | 0.03 | 0.58 | 0.03 | 0.93 |

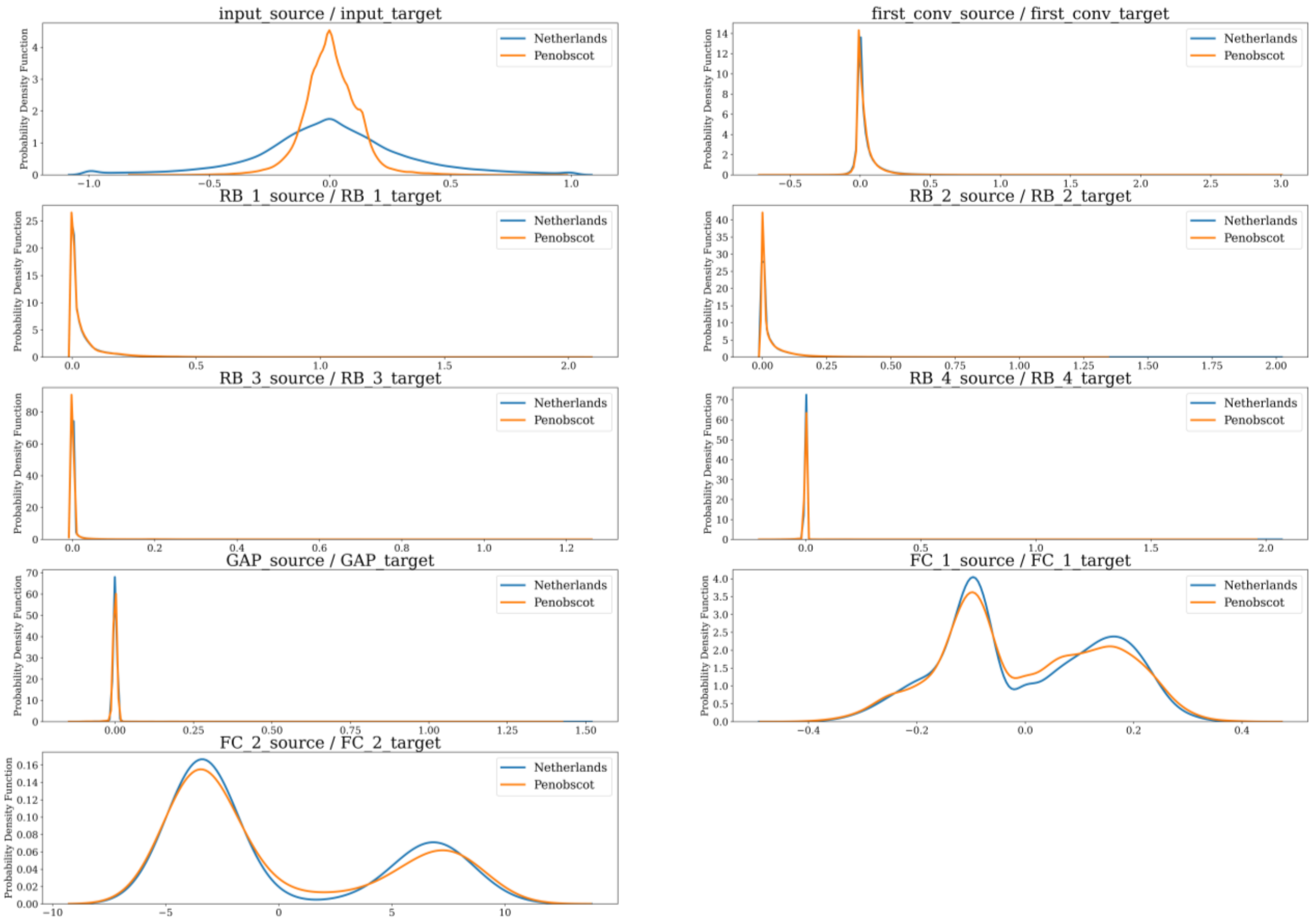

Supplementary Fig. 1. Comparing feature maps distributions of the source domain (SD) and target domain (TD). Distributions obtained from the architecture with 4 Encoders followed by a GAP layer and 2 FC layers (*4RB+GAP+2FC)* shows how the SD (Netherlands) and TD (Canada) distributions correlate.

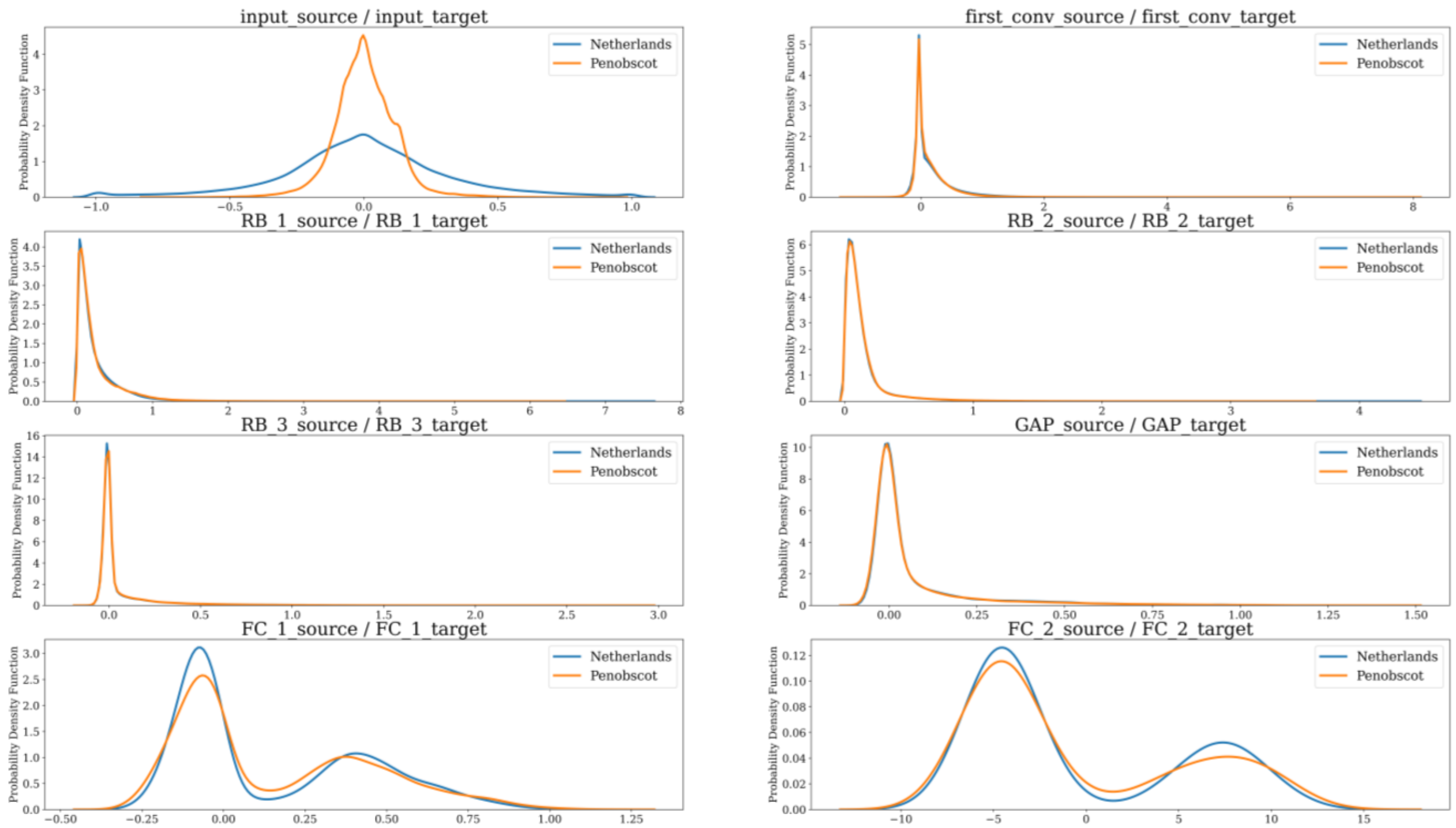

Supplementary Fig. 2. Comparing feature maps distributions of the source domain (SD) and target domain (TD). Distributions obtained from the architecture with 3 Encoders followed by a GAP layer and 2 FC layers (*3RB+GAP+2FC)* shows how the SD (Netherlands) and TD (Canada) distributions correlate.

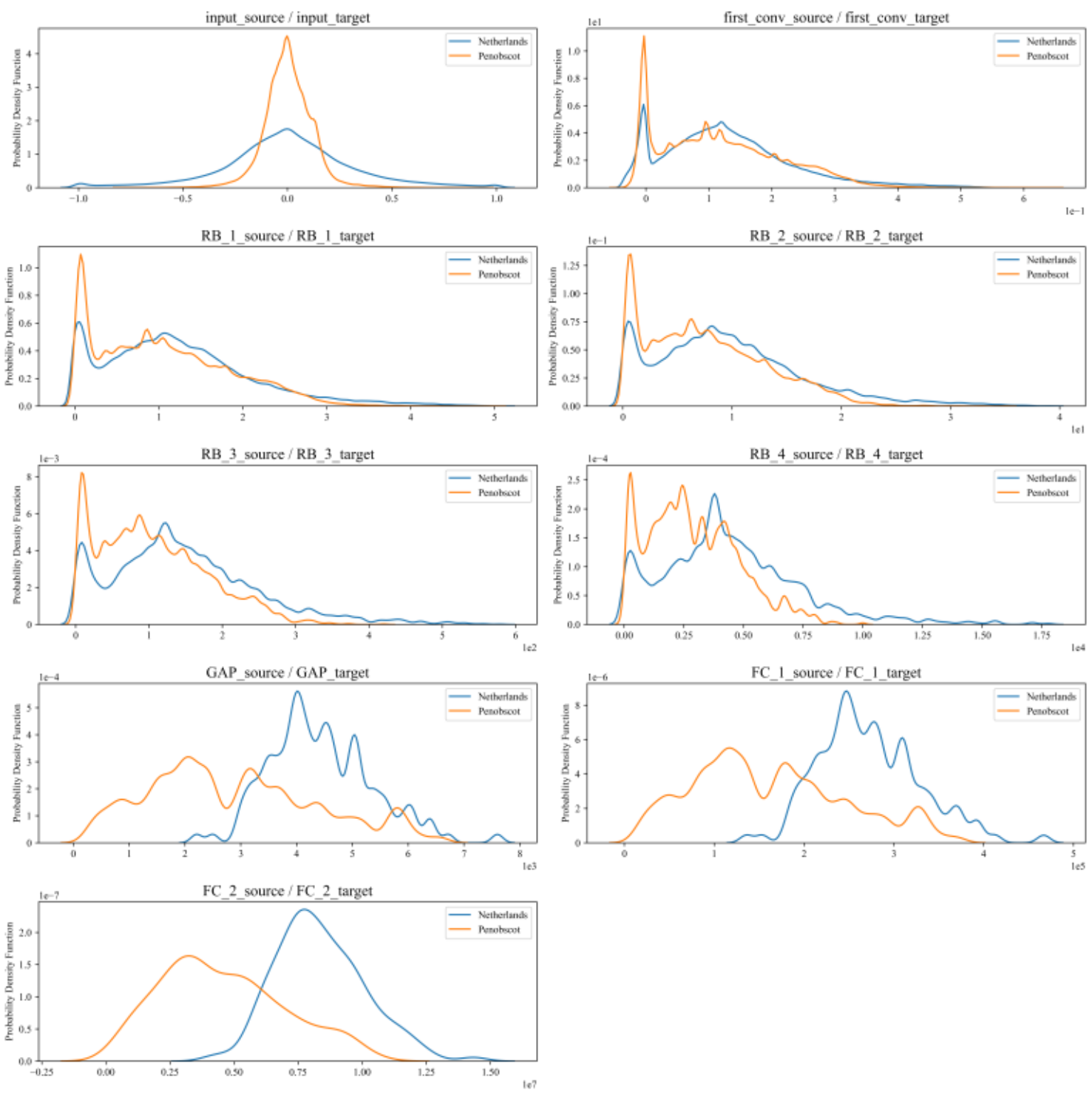

Supplementary Fig. 3. Comparison of the feature maps distributions from source and target domains when no CORAL Loss is applied, i.e., with random initialization. Distributions from the layer for the architecture with 4 Encoders followed by 4 FC layers (*4RB+GAP+2FC)* shows how the SD (Netherlands) and TD (Canada) distributions do not correlate when the model is not trained.

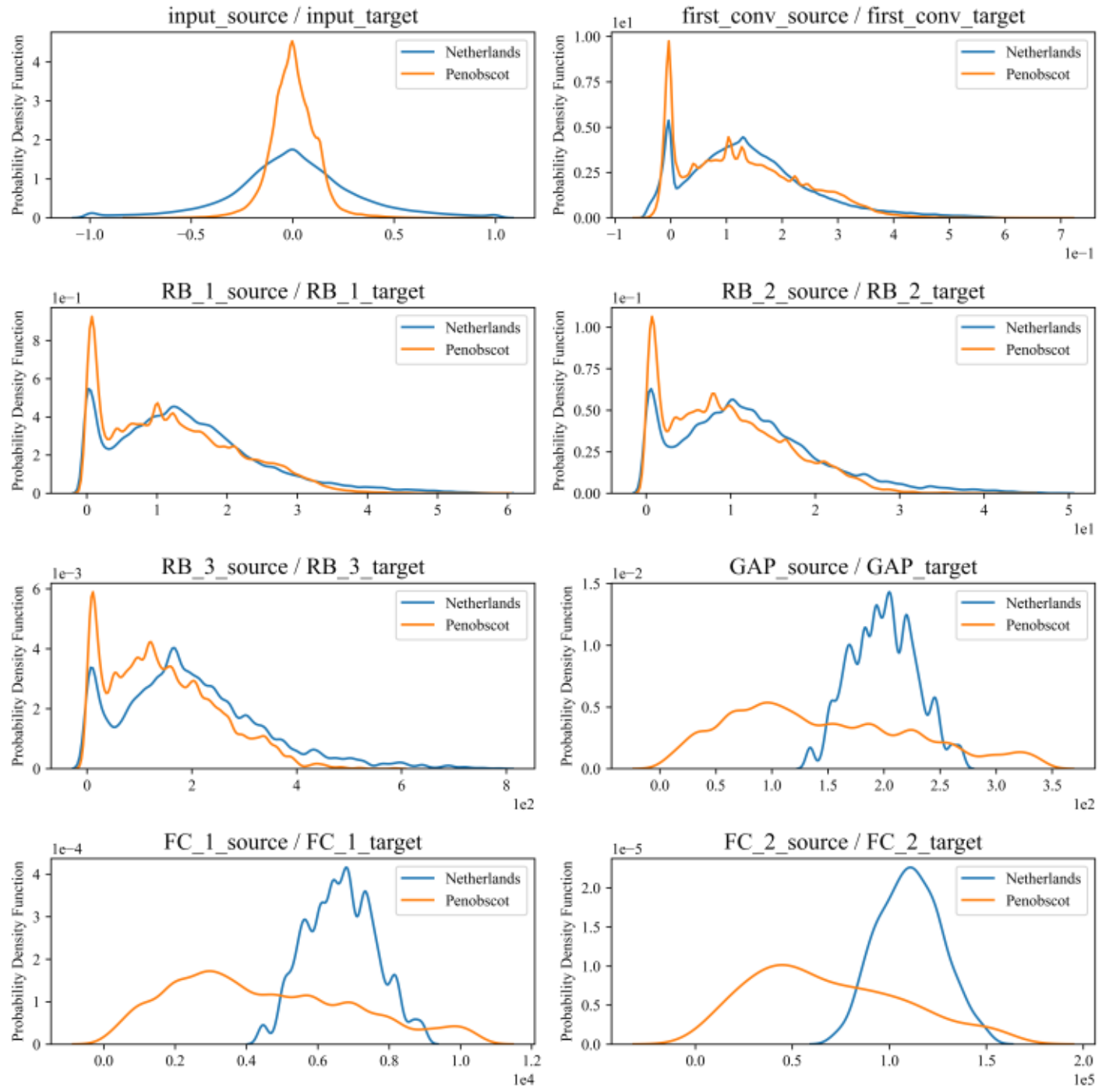

Supplementary Fig. 4. Comparison of the feature maps distributions from source and target domains when no CORAL Loss is applied, i.e., with random initialisation. Distributions from the layer for the architecture with 4 Encoders followed by 4 FC layers (*3RB+GAP+2FC)* shows how the SD (Netherlands) and TD (Canada) distributions do not correlate when the model is not trained.